%% file: main.tex
\documentclass[sigconf,screen,nonacm]{acmart}

\usepackage{multirow}
\usepackage{subcaption}
\AtBeginDocument{%
  \providecommand\BibTeX{{%
    \normalfont B\kern-0.5em{\scshape i\kern-0.25em b}\kern-0.8em\TeX}}}

\setcopyright{rightsretained}
\copyrightyear{2018}
\acmYear{2018}
\acmDOI{XXXXXXX.XXXXXXX}

\acmConference[Conference acronym 'XX]{Make sure to enter the correct
  conference title from your rights confirmation emai}{June 03--05,
  2018}{Woodstock, NY}
\acmPrice{15.00}
\acmISBN{978-1-4503-XXXX-X/18/06}

\begin{document}

\title{Digital Asset Valuation: A Study on Domain Names, Email Addresses, and NFTs}

\newcommand{\todo}[1]{\textsf{\textbf{\textcolor{cyan}{[[#1]]}}}}
\newcommand{\eg}{{e.g.}}
\newcommand{\ie}{{i.e.}}
\newcommand{\resp}{{resp. }}
\newcommand{\vs}{{vs. }}
\newcommand{\hhide}[1]{}
\newcommand{\dn}{{DASH}}
\newcommand{\dndomain}{{\dn\textsubscript{DN}}} %
\newcommand{\dnemail}{{\dn\textsubscript{EA}}} %
\newcommand{\dnnft}{{\dn\textsubscript{NFT}}}

\author{Kai Sun}
\authornote{Independent research.\\© 2022. All rights reserved.}
\email{kaisun.research@gmail.com}
\affiliation{%
  \institution{\ }
  \country{\ }
}

\renewcommand{\shortauthors}{Sun}

\input{0_abstract}

\ccsdesc[500]{Computing methodologies~Neural networks}
\ccsdesc[500]{Computing methodologies~Machine learning approaches}
\ccsdesc[100]{Computing methodologies~Natural language processing}
\ccsdesc[500]{Applied computing~Economics}
\keywords{digital asset, valuation, domain name, email address, non-fungible token, transaction, dataset, machine learning, language model} %

%
%
%

%
%
%
\maketitle

\input{1_introduction}

\input{2_relatedwork}

\input{3_data}

\input{4_approaches}
\input{5_experiments}

\input{6_conclusion}
\bibliographystyle{ACM-Reference-Format}
\bibliography{sample-base}

%
%

\end{document}

%% file: 0_abstract.tex
\begin{abstract}

Existing works on valuing digital assets on the Internet typically focus on a single asset class. To promote the development of automated valuation techniques, preferably those that are generally applicable to multiple asset classes, we construct {\dn}, the first Digital Asset Sales History dataset that features multiple digital asset classes spanning from classical to blockchain-based ones. Consisting of $280$K transactions of domain names ({\dndomain}), email addresses ({\dnemail}), and non-fungible token (NFT)-based identifiers ({\dnnft}), such as Ethereum Name Service names, {\dn} advances the field in several aspects: the subsets {\dndomain}, {\dnemail}, and {\dnnft} are the largest freely accessible domain name transaction dataset, the only publicly available email address transaction dataset, and the first NFT transaction dataset that focuses on identifiers, respectively.

We build strong conventional feature-based models as the baselines for {\dn}. We next explore deep learning models based on fine-tuning pre-trained language models, which have not yet been explored for digital asset valuation in the previous literature. We find that the vanilla fine-tuned model already performs reasonably well, outperforming all but the best-performing baselines. We further propose improvements to make the model more aware of the time sensitivity of transactions and the popularity of assets. Experimental results show that our improved model consistently outperforms all the other models across all asset classes on {\dn}.

\end{abstract}

%% file: 1_introduction.tex
\section{Introduction}

Since its birth, the Internet has generated many digital assets, such as domain names, and works on their monetary appraisal date back to over two decades ago~\cite{nguyen2001cyberproperty}. Most research on automated digital asset valuation focuses on a single asset class~\cite{wu2009domain,wu2009domain2,tang2014general,dieterle2014hybrid,bikadi2017prediction,liu2019data,delibacs2019domain,nadini2021mapping,kapoor2022tweetboost,jain2022nft}. Existing valuation methods rely heavily on expert knowledge and asset- and market-specific feature-engineering, whose cost reduces the potential for broadly applying the methods. Moreover, for many digital asset classes, there are no common testbeds or even no freely accessible data for studying valuation techniques, which raises the difficulties in making direct comparisons between methods, further limiting the progress in this research area.

With the goal of advancing the development of automated valuation methods, preferably those that are broadly applicable to multiple digital asset classes, we construct {\dn}, a \textbf{D}igital \textbf{A}sset \textbf{S}ales \textbf{H}istory dataset containing transactions of multiple representative asset classes. Specifically, {\dn} consists of sales history of domain names ({\dndomain}), email addresses ({\dnemail}), and non-fungible token (NFT)-based identifiers ({\dnnft}), such as Ethereum Name Service (ENS) names (Section~\ref{sec:data}). Assets of the classes featured in {\dn} are challenging to assess due to their non-fungibility~\cite{visconti2020valuation}. The valuation methods for these asset classes can potentially benefit from being collectively studied as they share the inherent property of being in some form of unique identifier.

To establish baseline performance on {\dn}, we design a set of features that apply to all the studied asset classes and build three conventional feature-based regression models (Section~\ref{sec:nonneural}). We next explore approaches based on fine-tuning pre-trained language models (LMs) (Section~\ref{sec:neural}). We find that the vanilla fine-tuned model performs reasonably well: without leveraging any handcrafted features or explicit expert knowledge, it surpasses two out of three conventional models on the average performance over all subsets of {\dn}. We further propose two improvements: (i) make the model more aware of the time sensitivity of transactions using a two-stage fine-tuning approach; (ii) append to the input sequence external knowledge about the popularity of assets. Experiments demonstrate that our improved model substantially reduces the mean squared logarithmic error (MSLE) by $4.2\%$ on average on the test set of {\dn} compared to the best-performing conventional model (Section~\ref{sec:experiment}).

Our main contributions are as follows.
\begin{itemize}
    \item We introduce {\dn}, the first digital asset transaction dataset that features multiple asset classes spanning from classical to blockchain-based ones. To our knowledge, (i) {\dndomain} is the largest freely accessible domain name transaction dataset; (ii) {\dnemail} is the only publicly available email address transaction dataset; and (iii) {\dnnft} is the first NFT transaction dataset that focuses on identifiers. %
    \item We propose conventional feature-based models and deep learning models for {\dn}. In contrast to all previous works, we present the first study that leverages pre-trained language models for digital asset valuation and demonstrates that fine-tuning a pre-trained language model can deliver performance superior to conventional models. %
    \item We conduct a comprehensive ablation study and a detailed error analysis of the proposed models on the {\dn} dataset. We also discuss variants of our models and the limitation of the work. The dataset and code will be available at \url{https://dataset.org/dash/}.
\end{itemize}

%% file: 2_relatedwork.tex
\section{Related Work}

\begin{figure}[ht!]
    \centering
    \begin{subfigure}[t]{0.48\textwidth}
        \centering
        \includegraphics[width=\textwidth]{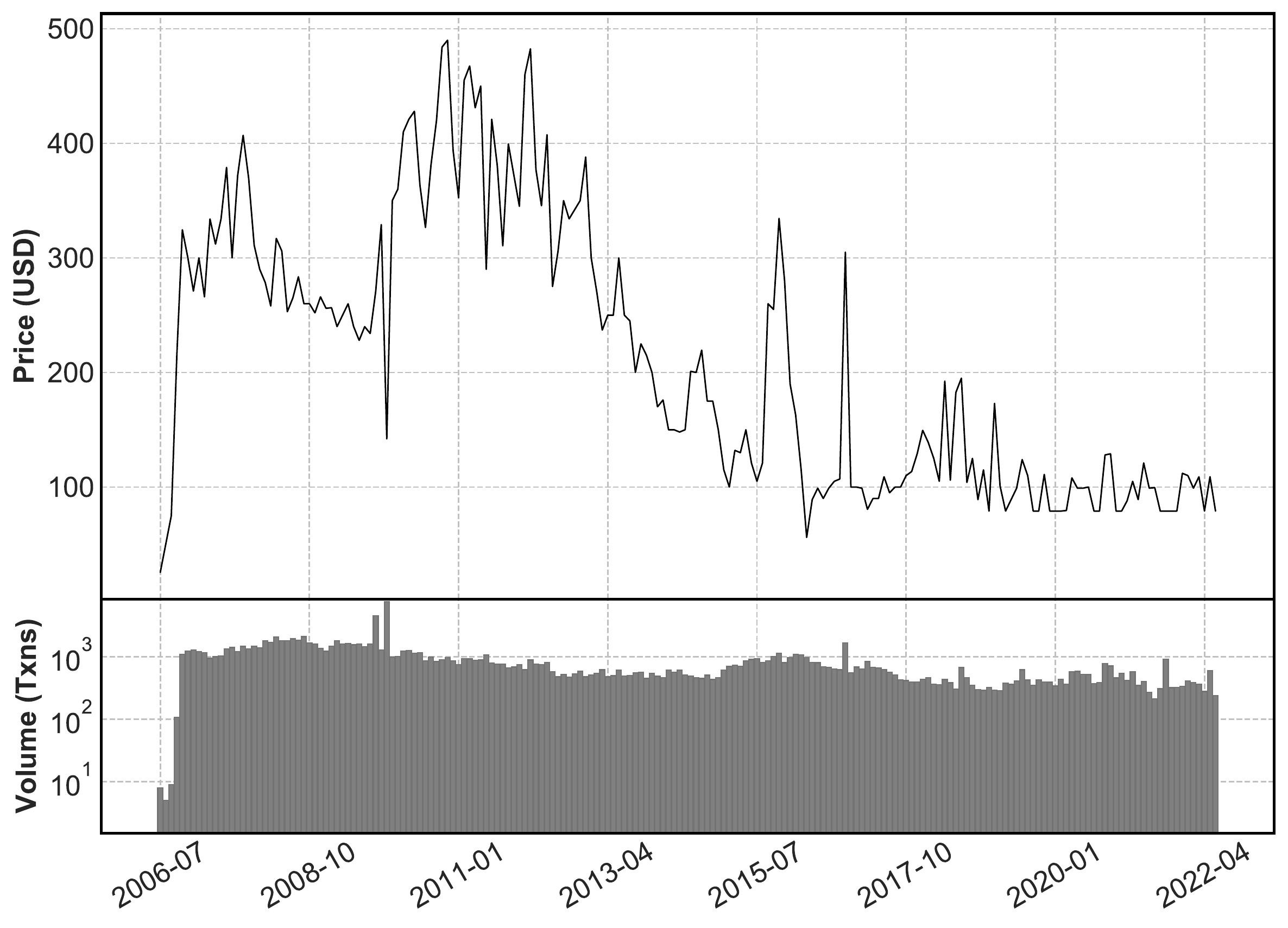}
        \caption{{\dndomain}.}
    \end{subfigure} %
    \begin{subfigure}[t]{0.48\textwidth}
        \centering
        \includegraphics[width=\textwidth]{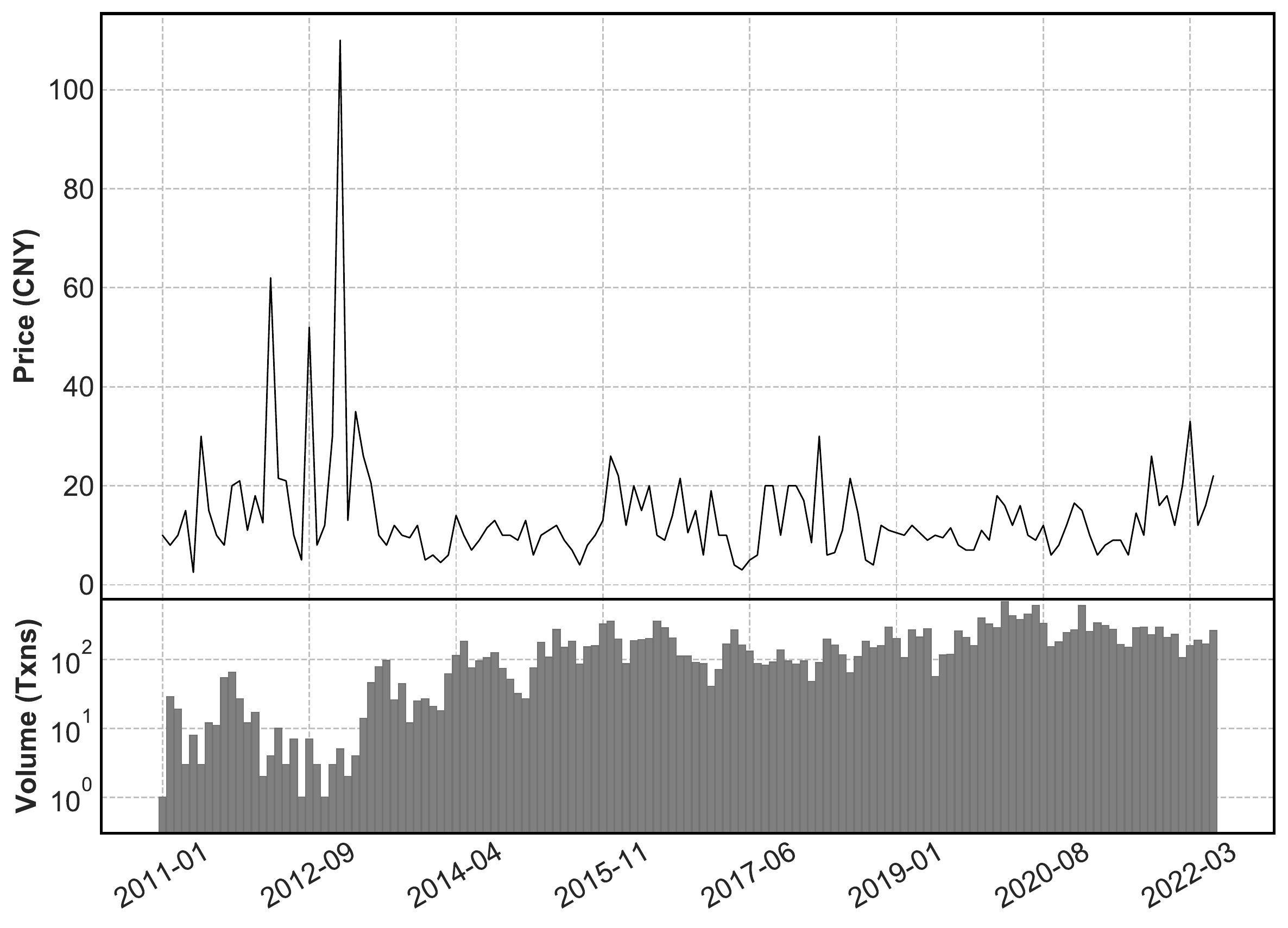}
        \caption{{\dnemail}.}
    \end{subfigure} %
    \begin{subfigure}[t]{0.48\textwidth}
        \centering
        \includegraphics[width=\textwidth]{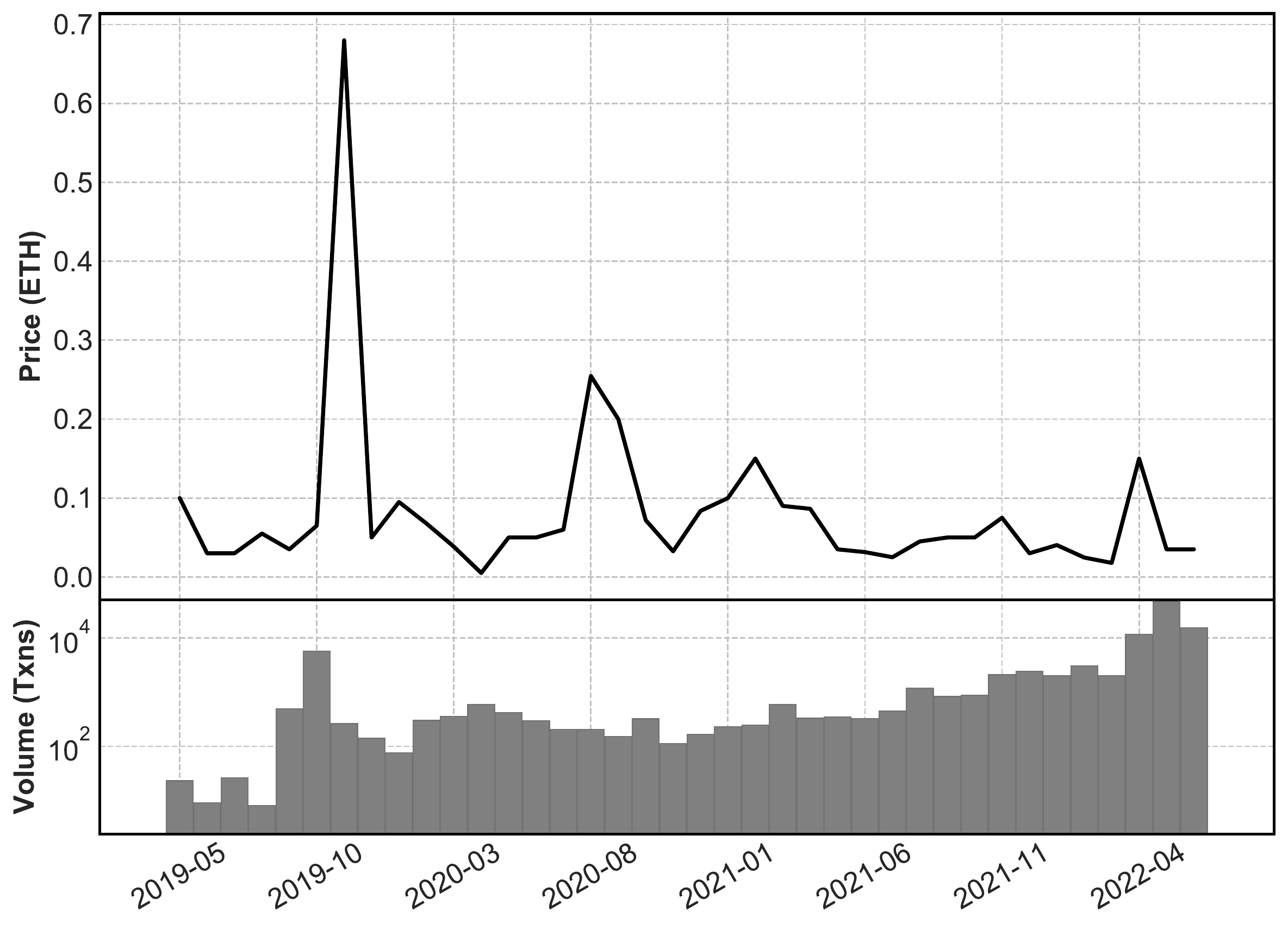}
        \caption{{\dnnft}.}
    \end{subfigure}
    \caption{Monthly transaction volume (in log scale) and median sales price (Txns: transactions).}
    \label{fig:datastat}
\end{figure}

\begin{table*}[ht!]
\centering
\begin{tabular}{lccc}
\toprule
& \bf {\dndomain} & \bf {\dnemail} & \bf {\dnnft} \\
\midrule
\# of transactions & $159,047$ & $20,308$  & $100,387$ \\
Avg./Max. \# of transactions per asset & $1.046$ / $17$ & $1.144$ / $8$ & $1.234$ / $13$ \\
Min./Med./Max. sales price (currency) & $1$ / $210$ / $5,000,000$ (USD) & $0.0001$ / $11$ / $11,111$ (CNY) & $1\times10^{-18}$ / $0.045$ / $420$ (ETH) \\
standard deviation of sales price & $25,492.0$ & $247.7$ & $2.3$ \\
Min./Med./Max. name length & $1$ / $9$ / $63$ & $1$ / $5$ / $21$ & $2$ / $5$ / $4,357$ \\
date range & 07/03/2006$\sim$06/30/2022 & 01/23/2011$\sim$06/30/2022 & 05/15/2019$\sim$06/30/2022 \\
\# of suffixes & $583$ & $642$ & $39$ \\
platforms (\% of transactions) & Sedo ($79.2\%$), Flippa ($20.8\%$) & FGLT ($100.0\%$) & OS ($95.5\%$), X2Y2 ($2.5\%$), LR ($2.1\%$) \\
\bottomrule
\end{tabular}
\caption{The overall statistics of {\dn} (OS: OpenSea, LR: LooksRare).} 
\label{tab:datastat}
\end{table*}

\subsection{Transaction Data}
\label{sec:related:data}
\subsubsection{Classical Assets}

We are not aware of any existing email address transaction datasets. As for domain name transactions, past studies use different data, most of which are not publicly accessible~\cite{wu2009domain2,dieterle2014hybrid,lindenthal2014valuable,bikadi2017prediction,delibacs2019domain}. The data released by \cite{liu2019data} is the only publicly accessible dataset we know. Their released dataset consists of $1,335$ domain names, which are all .com domains, along with binary labels indicating whether the price is high or low based on pre-defined thresholds. The exact sales price and date are not available. In comparison, {\dndomain} has orders of magnitude more sales with the exact sales price and date, covering over $500$ different domain extensions. Some platforms (\eg, NameBio\footnote{\url{https://namebio.com/}}) provide online domain name sales search services, but they do not support exporting data in batch freely.

\subsubsection{Blockchain-Based Assets}

Recent works have released several datasets regarding NFT transactions~\cite{nadini2021mapping,kapoor2022tweetboost,mekacher2022heterogeneous}. These datasets primarily focus on the NFTs' image objects ~\cite{nadini2021mapping,kapoor2022tweetboost}, traits assigned by the NFT creators~\cite{mekacher2022heterogeneous}, and mentions in social media~\cite{kapoor2022tweetboost}.

\subsection{Valuation Approaches}
\label{sec:related:approach}

We primarily discuss the works about valuing unique individual assets rather than predicting the price of fungible assets (\eg, the Bitcoin to USD price~\cite{dutta2020gated}) or estimating the aggregate price of asset collections (\eg, the median price of four-letter .com domain names, the average price of Bored Ape Yacht Club (BAYC)~\cite{jain2022nft}) over time. Various digital asset valuation methods, from theoretical to empirical, have been developed over the years~\cite{wu2009domain,wu2009domain2,tang2014general,dieterle2014hybrid,meystedt2015my,bikadi2017prediction,liu2019data,delibacs2019domain,visconti2020valuation,nadini2021mapping,kapoor2022tweetboost,mekacher2022heterogeneous}. Most methods employed in recent research are conventional feature-based machine learning models~\cite{liu2019data,delibacs2019domain,nadini2021mapping,kapoor2022tweetboost}. They show the superior performance of random forest, eXtreme Gradient Boosting (XGBoost)~\cite{chen2016xgboost}, and Adaptive Boosting (AdaBoost)~\cite{freund1995decision} for the valuation of domain names or NFTs compared with other feature-based models given the same feature set. A few works on NFT valuation leverage deep learning models, but the models are mainly used as the image encoder to encode image-based NFTs~\cite{nadini2021mapping,kapoor2022tweetboost}. Outside of academia, there are several well-established proprietary domain name appraisal systems from the industry, such as Estibot\footnote{\url{https://www.estibot.com/}} and GoDaddy Domain Appraisals\footnote{\url{https://www.godaddy.com/domain-value-appraisal}} (GoValue). In particular, GoValue employs deep learning to leverage the vast amount of domain name transaction data available only to GoDaddy\footnote{\url{https://www.godaddy.com/engineering/2019/07/26/domain-name-valuation/}}. Compared with all these works, we focus on methods generally applicable to multiple asset classes and demonstrate the effectiveness of models based on language model fine-tuning over previous methods in digital asset valuation.

\subsection{Fine-Tuning Pre-Trained Language Models}

The past few years have witnessed significant progress in various natural language understanding problems with the help of fine-tuning pre-trained high-capacity language models~\cite{radford2018improving,devlin2019bert}. More recently, applications of language model fine-tuning to problems beyond natural language understanding have emerged, such as automated theorem proving~\cite{polu2020generative} and playing chess~\cite{stockl2021watching}. We follow this thread and explore the application to digital asset valuation, which is underexplored in previous research.

%% file: 3_data.tex
\section{Data}
\label{sec:data}

\subsection{Collection Methodology}

\subsubsection{Data Sources}
We collect digital asset transaction data from a variety of data sources, summarized in the following.
\begin{itemize}
    \item \textbf{Domain names:} We track the domain name auctions hosted by sedo.com (Sedo) and flippa.com (Flippa). Additionally, we track the publicly disclosed buy-it-now sales and sales by negotiation completed on Flippa.
    \item \textbf{Email addresses:} We track the email address auctions hosted by fglt.net (FGLT).
    \item \textbf{NFTs:} We track sales of NFT-based identifiers, including ENS, Unstoppable Domains, and Decentraland Names reported by opensea.io (OpenSea), x2y2.io (X2Y2), and looksrare.org (LooksRare).
\end{itemize}

\subsubsection{Filtering and Normalization}
We filter out bundle sales and sales whose price is zero. For auctions, we only keep the ``reserve met'' and ``no reserve'' ones that have at least one bid. As ENS names are stored in a hashed form~\cite{xia2021ethereum}, we filter out ENS name transactions whose unhashed name is unknown. We convert the sales price to the dominant currency used in transactions of the corresponding asset class (\ie, USD, CNY, and ETH for domain names, email addresses, and NFTs, respectively) using the exchange rate at the transaction time.  

\subsubsection{Suspicious Transaction Detection} 

We notice suspicious transactions in the collected data. For example, the ENS name \emph{oneboy.eth} has been traded between two Ethereum addresses over $800$ times, indicating that a single agent likely controls the two addresses and these sales are likely bogus. To reduce the potential negative impact of such transactions, we adopt the following strategies: (i) for each transaction $t$ of NFT $x$ from address $a$ to address $b$, we remove $t$ if both $a$ and $b$ are involved in (but not necessarily together) at least two other transactions of $x$; (ii) for each transaction $t$ of email address $x$, we remove $t$ if another transaction of $x$ happens after $t$ within seven days. We do not apply similar strategies to domain names because Sedo and Flippa have relatively high commission fees, making it expensive to initiate suspicious transactions.

The resulting transactions of domain names, email addresses, and NFTs constitute {\dndomain}, {\dnemail}, and {\dnnft}, respectively, and $\text{{\dn}}=\text{{\dndomain}}\cup\text{{\dnemail}}\cup\text{{\dnnft}}$. Each transaction in {\dn} comprises an asset identifier, a transaction date, and the corresponding sales price, along with the meta information of the asset that consists of the asset class and asset collection (if applicable). Note that the studied assets in {\dn} do not have name collisions so far, so an asset identifier alone can unambiguously represent an asset. The meta information offers additional information to distinguish between different assets that share the same identifier in case there are name collisions in the future.

\subsection{Data Statistics}

For convenience, we first formally define the \emph{name} and \emph{suffix} of an asset, which will be referred to in the rest of this paper: given an asset from {\dn}, its name refers to the substring of the identifier starting from the beginning to the first delimiter (exclusive), and its suffix refers to the substring after the first delimiter (exclusive). The delimiter is at sign for {\dnemail} and dot for {\dndomain} and {\dnnft}. For example, the name and suffix of \emph{example.eth} are \emph{example} and \emph{eth}, respectively; the name and suffix of \emph{email@example.com} are \emph{email} and \emph{example.com}, respectively.

We summarize the statistics of {\dn} in Table~\ref{tab:datastat}. We see a considerable sales price range with a large standard deviation in every subset of {\dn}. Besides, most assets in {\dn} have only one transaction, and there are numerous different suffixes. These altogether provide evidence that {\dn} is a very challenging dataset for digital asset valuation. We plot Figure~\ref{fig:datastat} to show an overall view of transaction distributions across time. Compared with {\dndomain} and {\dnemail}, the transactions in {\dnnft} are less evenly distributed in terms of volume over time. %

%% file: 4_approaches.tex
\section{Approaches}
\label{sec:method}

\subsection{Task Formulations}

In this work, we enforce that transactions in the development set are newer (\resp older) compared with the training (\resp test) set. This is distinguished from many previous works, where training data can be newer than evaluation data~\cite{delibacs2019domain,kapoor2022tweetboost}. Specifically, we split the data by date, with the latest $5\%$ for testing, the next latest $5\%$ for development, and the earliest $90\%$ for training. We summarize the split in Table~\ref{tab:split}. As the date range in the split is different for different asset classes, to prevent time leakage, we train separate models for {\dndomain}, {\dnemail}, and {\dnnft}.

\begin{table}[t!]
\centering
\begin{tabular}{llccc}
\toprule
& & \bf Date Range & \bf \# of Txns \\ %
\midrule
& Train & 07/03/2006 $\sim$ 07/06/2019 & $143,147$ \\
{\dndomain} & Dev & 07/07/2019 $\sim$ 11/19/2020 & $7,973$ \\
& Test & 11/20/2020 $\sim$ 06/30/2022 & $7,927$ \\
\midrule
& Train & 01/23/2011 $\sim$ 09/11/2021 & $18,288$ \\
{\dnemail} & Dev & 09/12/2021 $\sim$ 01/13/2022 & $1,006$ \\
& Test & 01/14/2022 $\sim$ 06/30/2022 & $1,014$ \\
\midrule
& Train & 05/15/2019 $\sim$ 06/18/2022 & $90,781$ \\
{\dnnft} & Dev & 06/19/2022 $\sim$ 06/27/2022 & $4,882$ \\
& Test & 06/28/2022 $\sim$ 06/30/2022 & $4,724$ \\
\bottomrule
\end{tabular}
\caption{Data splitting (Txns: transactions).} 
\label{tab:split}
\end{table}

For consistency and clarity, we now introduce several notations and give a formal definition of the task. Given a set of transactions $S=\left\{(a_1,p_1,t_1),(a_2,p_2,t_2),\ldots\right\}$ where $a_i$, $p_i$, and $t_i$ represent the asset identifier, price, and time of the $i^{\text{th}}$ transaction, respectively, and $S\in\left\{\text{\dndomain}, \text{\dnemail}, \text{\dnnft}\right\}$, we partition $S$ into $S_\text{train}$, $S_\text{dev}$, and $S_\text{test}$ such that $\forall (a_i,p_i,t_i)\in S_\text{train}, (a_j,p_j,t_j)\in S_\text{dev}, (a_k,p_k,t_k)\in S_\text{test} (t_i\leq t_j\leq t_k)$. The task is to learn from $S_\text{train}$ a model that takes as input an asset identifier, and outputs an estimation of the price as accurate as possible, measured by MSLE on $S_\text{dev}$ and $S_\text{test}$.

\begin{table*}[ht!]
\centering
\begin{tabular}{lcccccccc}
\toprule
\multirow{2}{*}{\textbf{Method}} & \multicolumn{2}{c}{\textbf{\dndomain}} & \multicolumn{2}{c}{\textbf{\dnemail}} & \multicolumn{2}{c}{\textbf{\dnnft}} & \multicolumn{2}{c}{\textbf{Average}} \\
& \bf Dev & \bf Test & \bf Dev & \bf Test & \bf Dev & \bf Test & \bf Dev & \bf Test \\
\midrule
Mean & $3.248$ & $2.748$ & $2.809$ & $2.713$ & $2.640$ & $2.567$ & $2.899$ & $2.676$ \\
\midrule
AdaBoost & $2.382$ & $1.999$ & $1.838$ & $2.064$ & $1.524$ & $1.530$ & $1.915$ & $1.864$  \\
Random Forest & $2.518$ & $2.187$ & $1.393$ & $1.594$ & $1.190$ & $1.149$ & $1.700$ & $1.643$ \\
XGBoost & $2.218$ & $1.940$ & $1.253$ & $1.497$ & $1.081$ & $1.077$ & $1.517$ & $1.505$ \\
\midrule
Vanilla mBERT & $2.495$ & $2.140$ & $1.375$ & $1.400$ & $1.111$ & $1.312$ & $1.660$ & $1.617$ \\
mBERT$+$ & $\mathbf{2.106}$ & $\mathbf{1.892}$ & $\mathbf{1.212}$ & $\mathbf{1.368}$ & $\mathbf{1.007}$ & $\mathbf{1.066}$ & $\mathbf{1.442}$ & $\mathbf{1.442}$ \\
\bottomrule
\end{tabular}
\caption{Performance in MSLE.} %
\label{tab:main}
\end{table*}

\begin{table*}[ht!]
\centering
\begin{tabular}{lcccccccc}
\toprule
\multirow{2}{*}{\textbf{Method}} & \multicolumn{2}{c}{\textbf{\dndomain}} & \multicolumn{2}{c}{\textbf{\dnemail}} & \multicolumn{2}{c}{\textbf{\dnnft}} & \multicolumn{2}{c}{\textbf{Average}} \\
& \bf MSLE & $\mathbf{\Delta}$ & \bf MSLE & $\mathbf{\Delta}$ & \bf MSLE & $\mathbf{\Delta}$ & \bf MSLE & $\mathbf{\Delta}$ \\
\midrule
XGBoost & $2.218$ & -- & $1.253$ & -- & $1.081$ & -- & $1.517$ & -- \\
-- length & $2.244$ & $+0.026$ & $1.431$ & $+0.178$ & $1.302$ & $+0.221$ & $1.659$ & $+0.142$ \\
-- suffix & $2.645$ & $+0.427$ & $1.834$ & $+0.581$ & $1.154$ & $+0.073$ & $1.878$ & $+0.361$ \\
-- character & $2.223$ & $+0.005$ & $1.424$ & $+0.171$ & $1.173$ & $+0.092$ & $1.607$ & $+0.090$ \\
-- \# of tokens & $2.227$ & $+0.009$ & $1.280$ & $+0.027$ & $1.089$ & $+0.008$ & $1.532$ & $+0.015$ \\
-- vocabulary & $2.225$ & $+0.007$ & $1.293$ & $+0.040$ & $1.093$ & $+0.012$ & $1.537$ & $+0.020$ \\
-- trademark & $2.231$ & $+0.013$ & $1.297$ & $+0.044$ & $1.091$ & $+0.010$ & $1.540$ & $+0.023$ \\
-- TLD count & $2.548$ & $+0.330$ & $1.449$ & $+0.196$ & $1.151$ & $+0.070$ & $1.716$ & $+0.199$ \\
\midrule
mBERT$+$ & $2.106$ & -- & $1.212$ & -- & $1.007$ & -- & $1.442$ \\
-- LM pre-training & $2.203$ & $+0.097$ & $1.570$ & $+0.358$ & $1.173$ & $+0.166$ & $1.649$ & $+0.207$ \\
-- 1st stage fine-tuning & $2.411$ & $+0.305$ & $1.422$ & $+0.210$ & $1.165$ & $+0.158$ & $1.666$ & $+0.224$ \\
-- 2nd stage fine-tuning & $2.438$ & $+0.332$ & $1.215$ & $+0.003$ & $1.107$ & $+0.100$ & $1.587$ & $+0.145$ \\
-- TLD count & $2.396$ & $+0.290$ & $1.382$ & $+0.170$ & $1.064$ & $+0.057$ & $1.614$ & $+0.172$ \\
\bottomrule
\end{tabular}
\caption{Ablation tests on the development set.} 
\label{tab:ablation}
\end{table*}
\subsection{Non-Neural Models}
\label{sec:nonneural}
\subsubsection{Mean Value Baseline}

This simple model predicts a constant value that minimizes the MSLE on the training set. More formally, the constant value is the geometric mean price $\tilde{p}$ defined as
$$
\tilde{p}=\exp\left(\frac{1}{|S_\text{train}|}\sum_{(a_i,p_i,t_i)\in S_\text{train}} \log(p_i)\right)\text{.}
$$
\subsubsection{Feature-Based Regression Models}

Following previous empirical studies (Section~\ref{sec:related:approach}), we develop conventional feature-based models using random forest, XGBoost, and AdaBoost. Inspired by previous studies on drop catching and squatting~\cite{miramirkhani2018panning,xia2021ethereum}, we design the following feature set reflecting the intrinsic value of identifiers, which applies to all the asset classes studied in this work.

\begin{itemize}
    \item \textbf{Length:} the length of the asset name in character.
    \item \textbf{Suffix:} the asset suffix, represented by a one-hot vector.
    \item \textbf{Character:} four binary features indicating (i) whether the asset name only contains alphabet letters, (ii) whether the asset name contains hyphens, (iii), whether all the characters in the asset name are numeric, and (iv) whether the asset name contains non-ASCII characters, respectively.
    \item \textbf{Number of tokens:} we tokenize the asset name using morphological analysis~\cite{virpioja2013morfessor} and consider the number of tokens as a feature. 
    \item \textbf{Vocabulary:} two binary features indicating (i) whether the asset name is a word and (ii) whether the asset name is an adult word, respectively.
    \item \textbf{Trademark:} a binary feature indicating if the asset name appears in any trademark applications.%
    \item \textbf{Top-level domain (TLD) count:} the number of TLDs where the asset name is registered.
\end{itemize}

\subsection{Neural Models}
\label{sec:neural}
\subsubsection{Vanilla mBERT}

We follow the framework of fine-tuning a pre-trained high-capacity language model~\cite{radford2018improving} and use multilingual BERT (mBERT)~\cite{devlin2019bert} as the pre-trained model. Given an asset, we concatenate the name $n$ and suffix $s$ of the asset with the classification token \texttt{[CLS]} and separator token \texttt{[SEP]} in mBERT as the input sequence \texttt{[CLS]}$n$\texttt{[SEP]}$s$\texttt{[SEP]} to mBERT, with a linear layer on top of the final hidden state for \texttt{[CLS]} in the input sequence. %

\subsubsection{mBERT$+$}
We improve vanilla mBERT in two effective yet easy-to-implement ways:
\begin{itemize}
    \item[(i)] Since transaction data is time sensitive, we propose a two-stage fine-tuning approach to highlight the relatively new transactions during training. Specifically, in the first stage, we first fine-tune the pre-trained mBERT on all transactions in $S_\text{train}$. We then fine-tune the resulting model again in the second stage on the newest $T$ transactions in $S_\text{train}$.
    \item[(ii)] We propose a modification to the input sequence of Vanilla mBERT to leverage external knowledge that helps approximate the popularity of the asset name but is not readily available in the pre-trained representation. Concretely, the modified input sequence is \texttt{[CLS]}$n$\texttt{[SEP]}$s$\texttt{[SEP]}$c$\texttt{[SEP]}, where $c$ is a string of digits representing the TLD count (defined in Section~\ref{sec:nonneural}). 

\end{itemize}

%% file: 5_experiments.tex
\section{Experiments and Discussions}
\label{sec:experiment}
\subsection{Implementation Details}

\subsubsection{Feature Extraction}

We employ Polyglot\footnote{\url{https://github.com/aboSamoor/polyglot}} for morphological analysis. For vocabulary feature extraction, we use the vocabulary of GloVe.840B.300D~\cite{pennington2014glove} and a self-collected adult word list\footnote{We will release the adult word list along with the code.}. We look up trademark applications from April 1884 to December 2018 released by the United States Patent and Trademark Office. We leverage the DNS Census 2013~\footnote{\url{https://archive.org/details/DNSCensus2013}} to obtain TLD count.

\subsubsection{Non-Neural Models}

We implement random forest and AdaBoost using~\cite{pedregosa2011scikit} and XGBoost using~\cite{chen2016xgboost}. All hyperparameters take the default values.

\subsubsection{Neural Models}

We implement vanilla mBERT and mBERT$+$ based on Transformers~\cite{wolf2020transformers}. We use the multilingual uncased BERT-Base released by~\cite{devlin2019bert}. For vanilla mBERT, we fine-tune for three epochs. As for mBERT$+$, we set $T$ to $3,000$ and fine-tune for one epoch in the first fine-tuning stage and three epochs in the second fine-tuning stage. We set the learning rate and batch size to $2\times 10^{-5}$ and $64$, respectively. All unspecified hyperparameters take the default values~\cite{devlin2019bert}.

\subsection{Main Results}

We report in Table~\ref{tab:main} the performance of all models introduced in Section~\ref{sec:method}. XGBoost consistently performs the best among conventional feature-based models across all subsets of {\dn}. Vanilla mBERT, which does not leverage any handcrafted features or explicit expert knowledge, outperforms AdaBoost and random forest in average performance, showing the potential of language model fine-tuning for digital asset valuation. We see a significant reduction in MSLE relative to vanilla mBERT when employing the proposed improvements. Compared with XGBoost, the improved model mBERT$+$ substantially reduces the MSLE by $4.2\%$ (\ie, $1.442$ \vs $1.505$) on average on the test set of {\dn} ($\text{p-value}<0.005$).

To measure the contribution of different components, we conduct ablation tests, where we remove one component from XGBoost or mBERT$+$ at a time. As shown in Table~\ref{tab:ablation}, the suffix, TLD count, and length features contribute the most to the performance of XGBoost. Every component of mBERT$+$ heavily impacts the overall performance. Specifically, compared with a single-stage fine-tuning on the entire (\resp the most recent portion of) training data, the two-stage fine-tuning reduces the MSLE by $0.145$ (\resp $0.224$) on average. Incorporating TLD count in the input sequence contributes to an average decrease of $0.172$ in MSLE. Furthermore, the MSLE increases by $0.207$ on average if we replace the pre-trained mBERT weights with randomly initialized weights.

\subsection{Error Analysis}
\label{sec:error}
\begin{figure}[t!]
    \centering
    \begin{subfigure}[t]{0.45\textwidth}
        \centering
        \includegraphics[width=\textwidth]{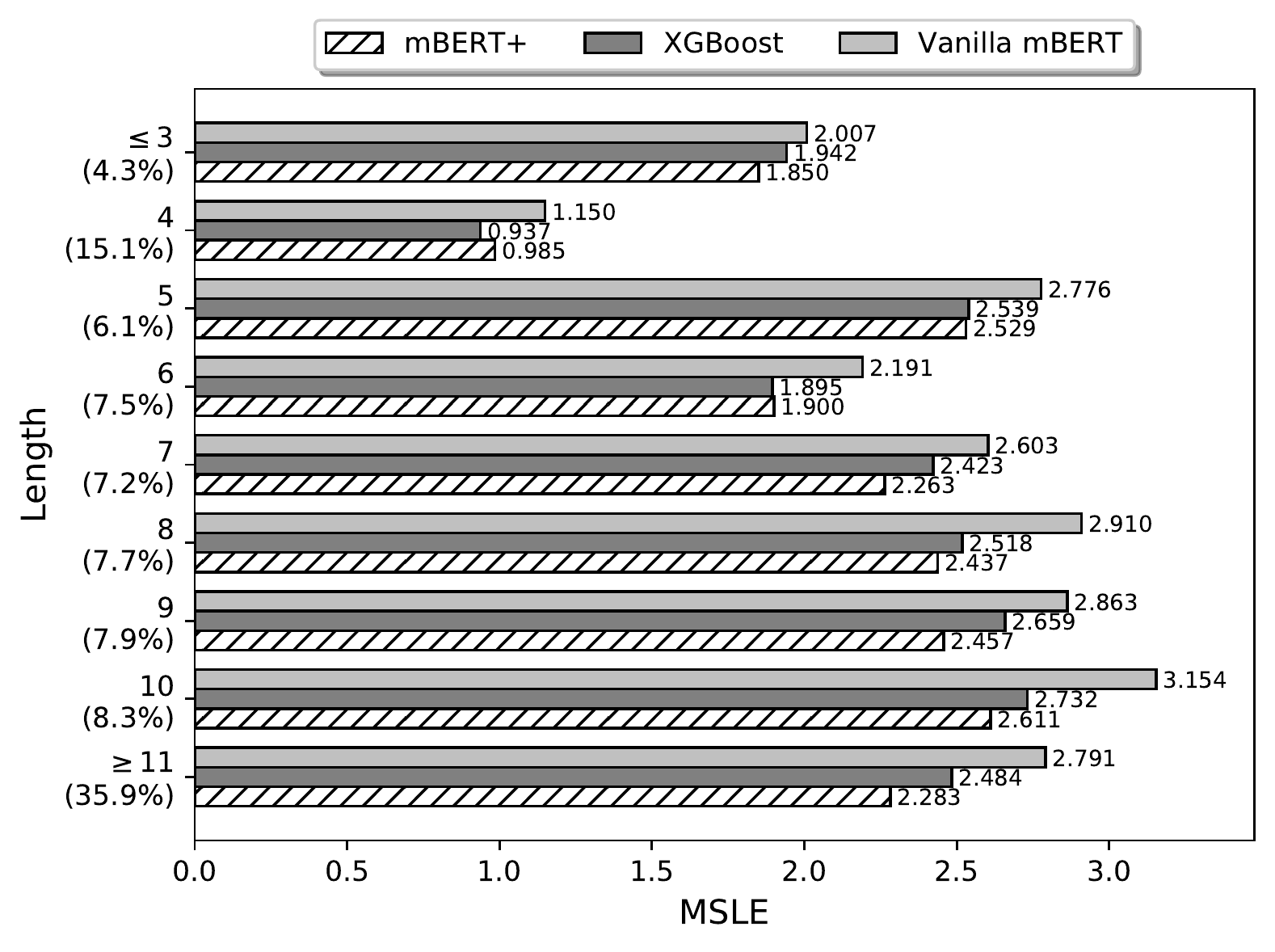}
        \caption{{\dndomain}.}
    \end{subfigure} %
    \begin{subfigure}[t]{0.45\textwidth}
        \centering
        \includegraphics[width=\textwidth]{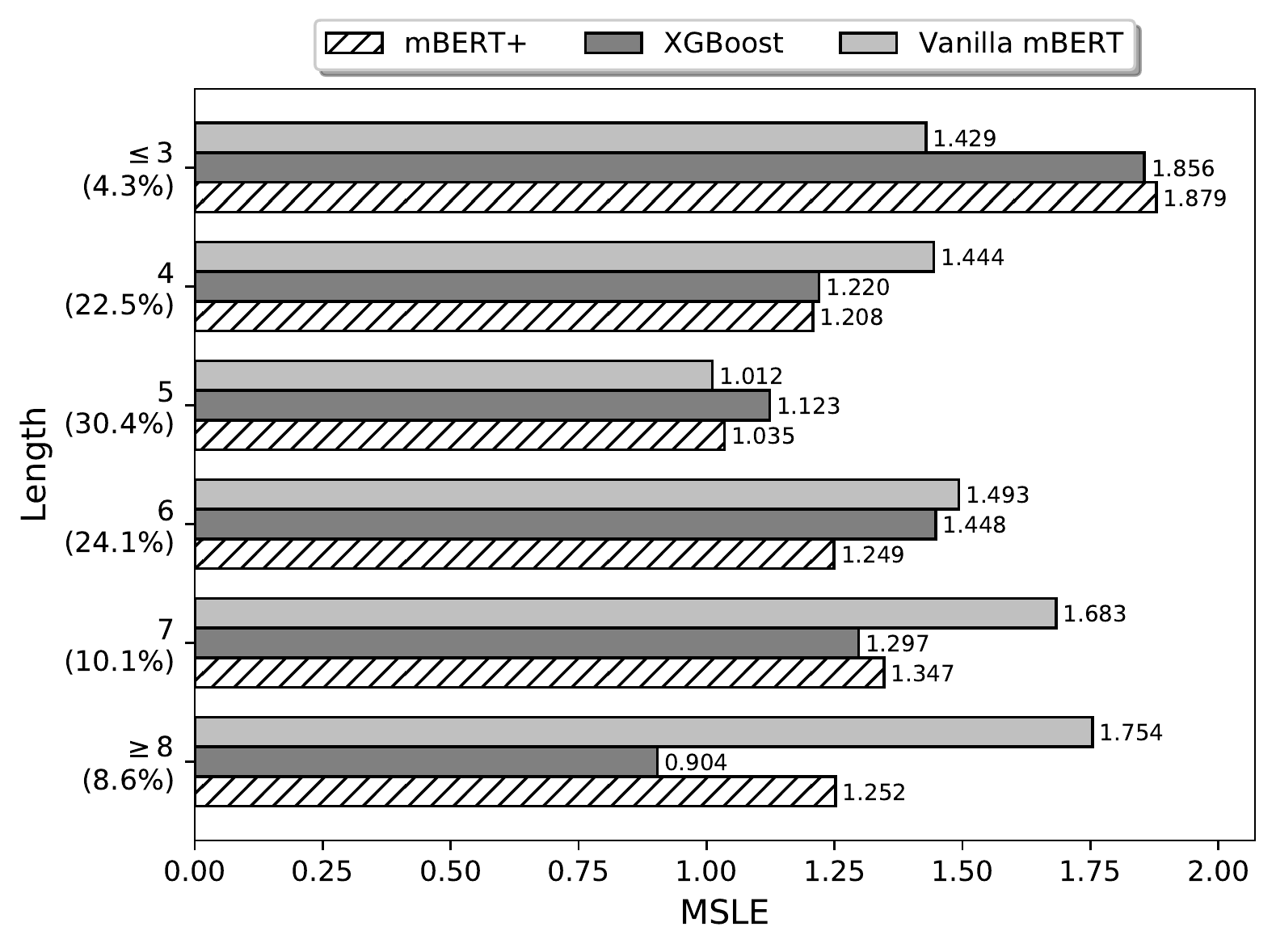}
        \caption{{\dnemail}.}
    \end{subfigure} %
    \begin{subfigure}[t]{0.45\textwidth}
        \centering
        \includegraphics[width=\textwidth]{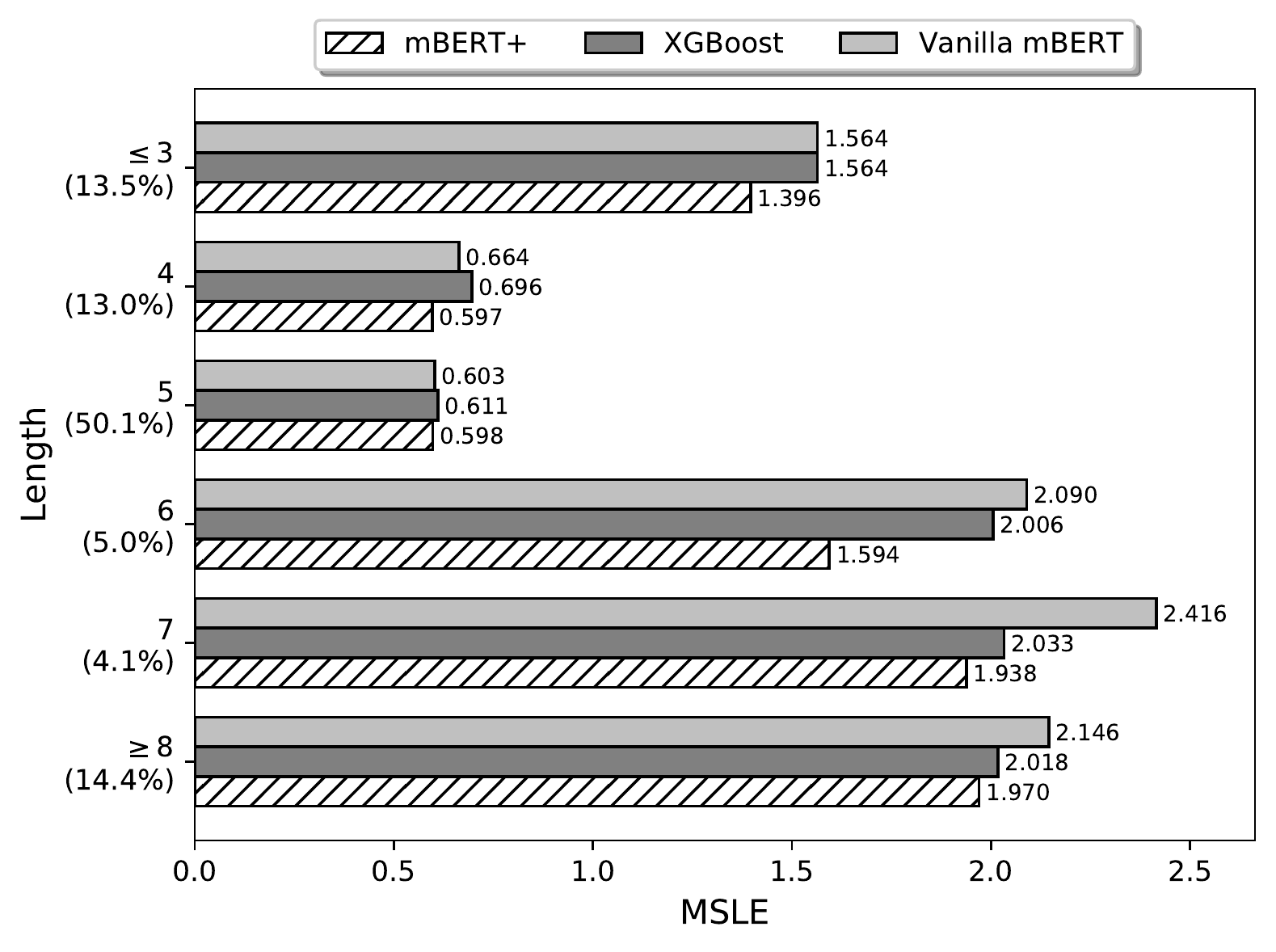}
        \caption{{\dnnft}.}
    \end{subfigure}
    \caption{Performance comparison in MSLE by name length on the development set. The percentage of each length group is in parentheses.}
    \label{fig:errorlength}
\end{figure}

\begin{figure}[t!]
    \centering
    \begin{subfigure}[t]{0.45\textwidth}
        \centering
        \includegraphics[width=\textwidth]{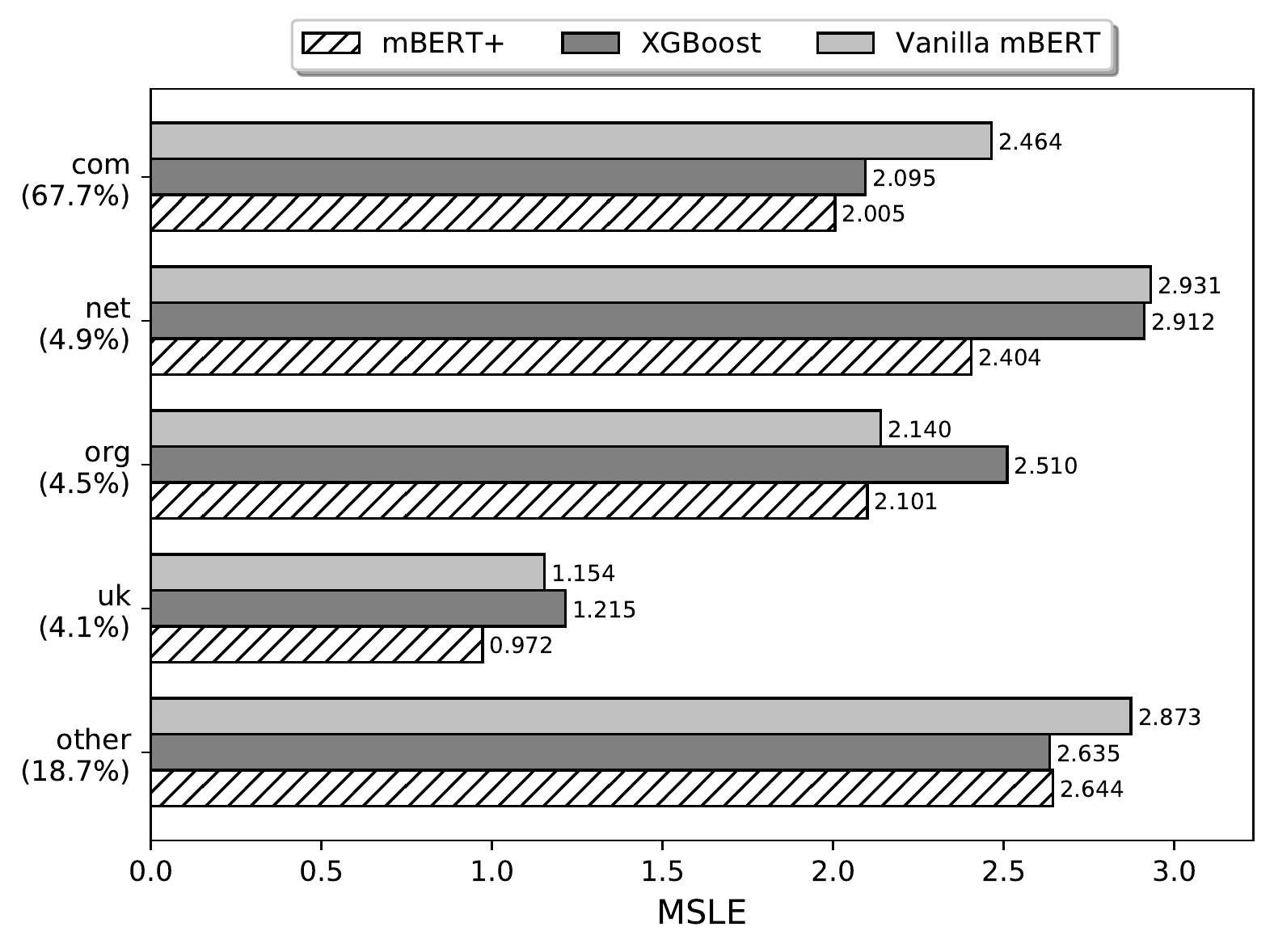}
        \caption{{\dndomain}.}
    \end{subfigure} %
    \begin{subfigure}[t]{0.45\textwidth}
        \centering
        \includegraphics[width=\textwidth]{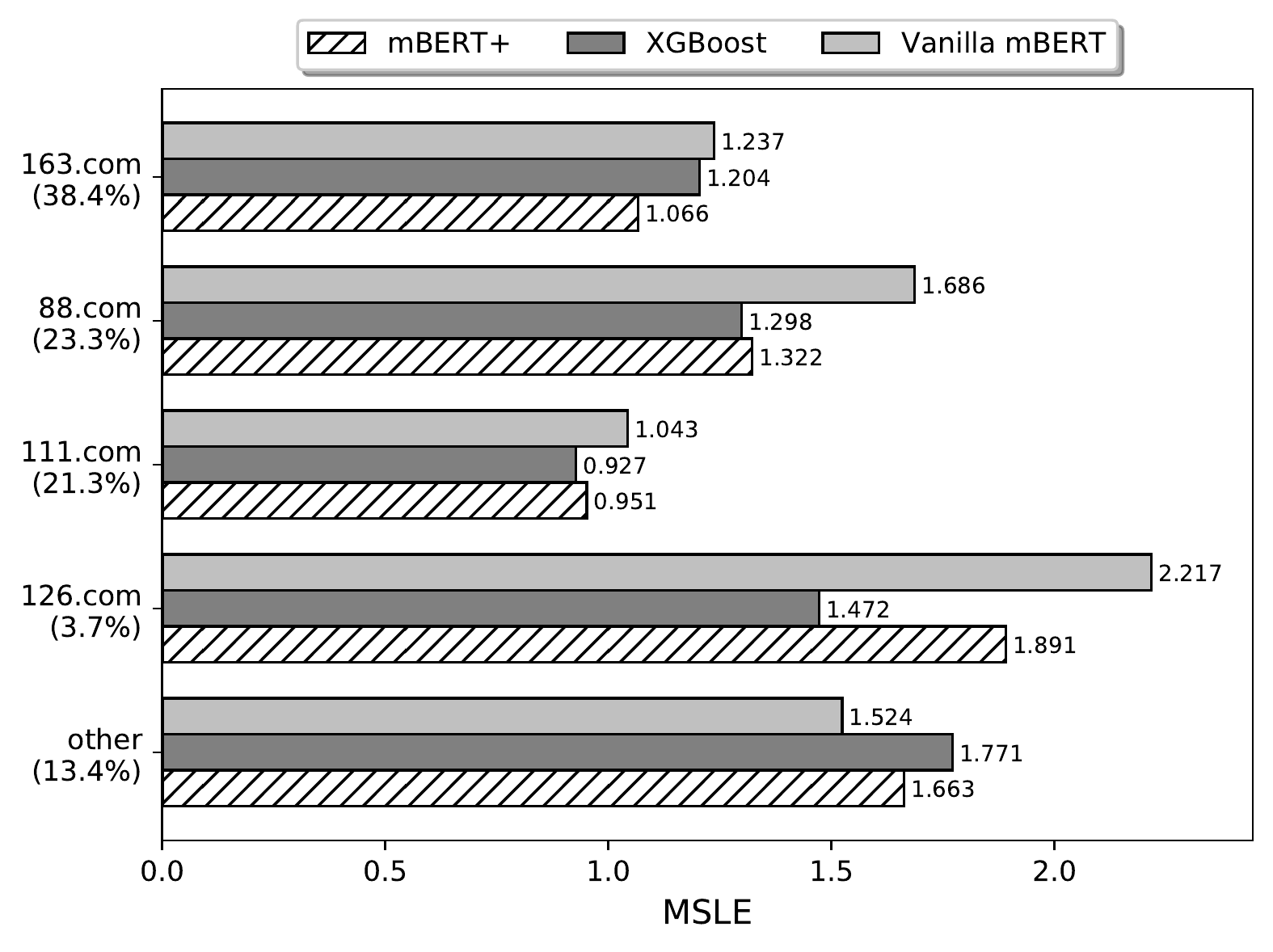}
        \caption{{\dnemail}.}
    \end{subfigure} %
    \begin{subfigure}[t]{0.45\textwidth}
        \centering
        \includegraphics[width=\textwidth]{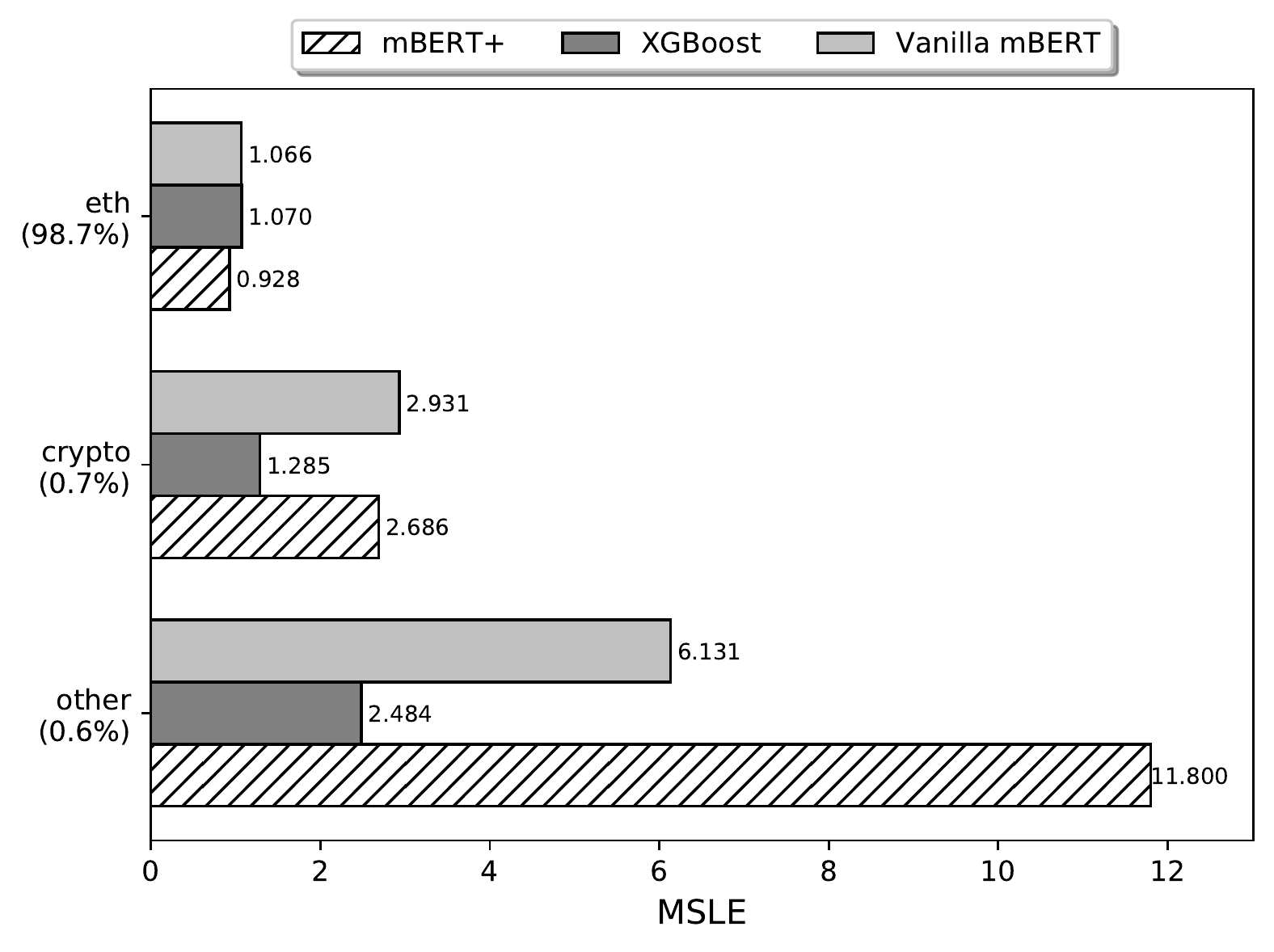}
        \caption{{\dnnft}.}
    \end{subfigure}
    \caption{Performance comparison in MSLE by suffix on the development set. The percentage of each suffix group is in parentheses.}
    \label{fig:errorsuffix}
\end{figure}

\begin{figure}[t!]
    \centering
    \begin{subfigure}[t]{0.45\textwidth}
        \centering
        \includegraphics[width=\textwidth]{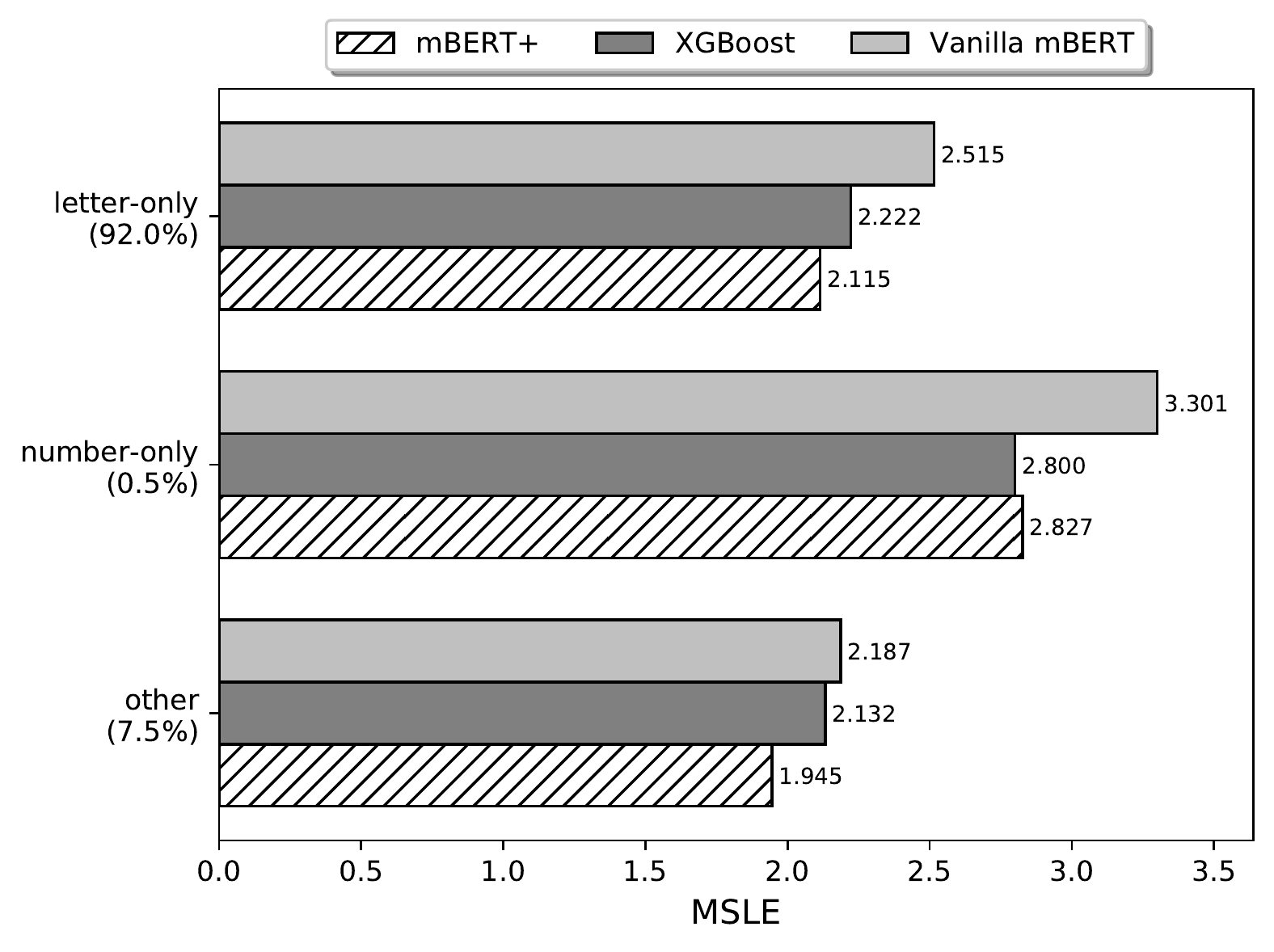}
        \caption{{\dndomain}.}
    \end{subfigure} %
    \begin{subfigure}[t]{0.45\textwidth}
        \centering
        \includegraphics[width=\textwidth]{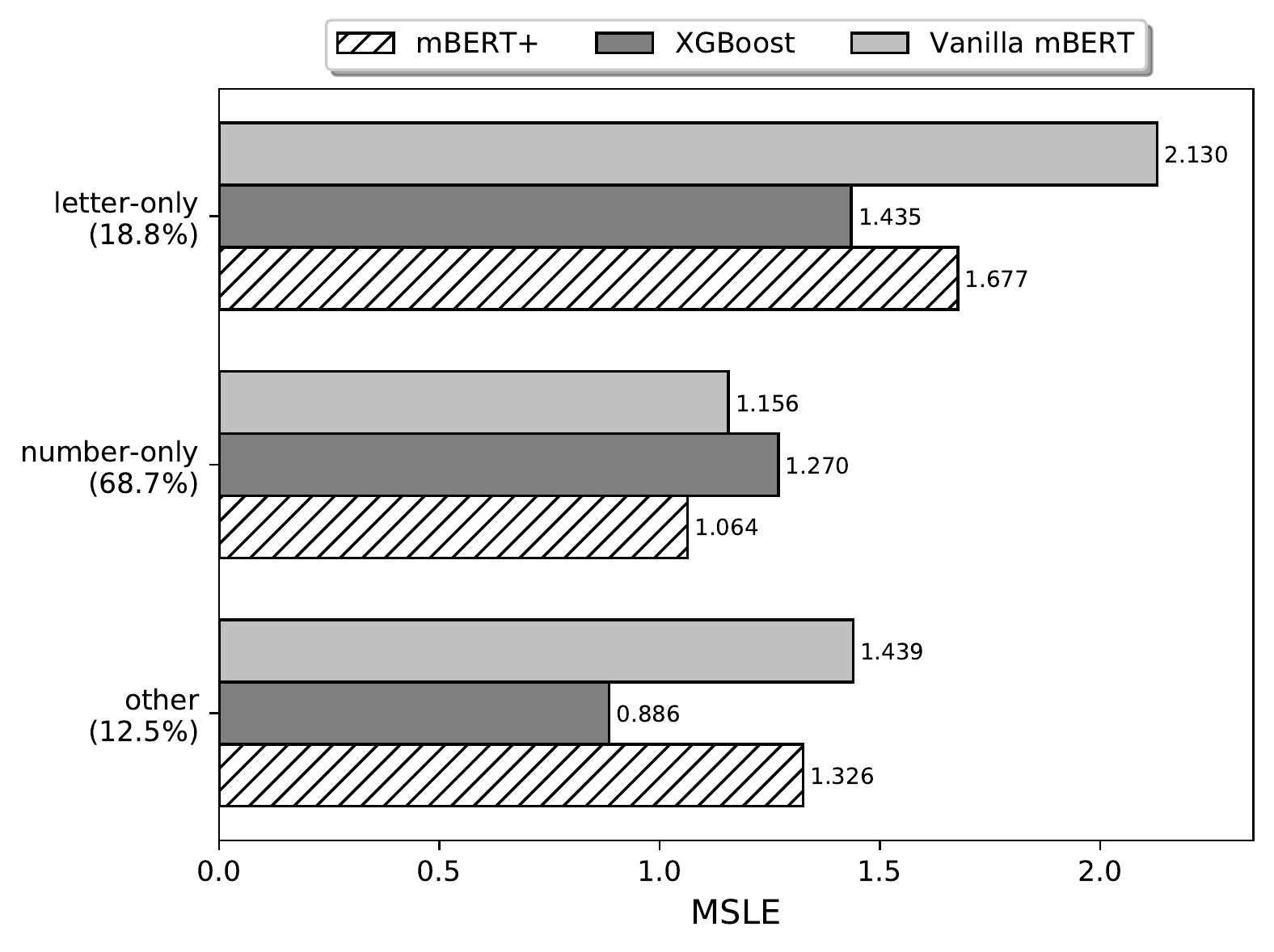}
        \caption{{\dnemail}.}
    \end{subfigure} %
    \begin{subfigure}[t]{0.45\textwidth}
        \centering
        \includegraphics[width=\textwidth]{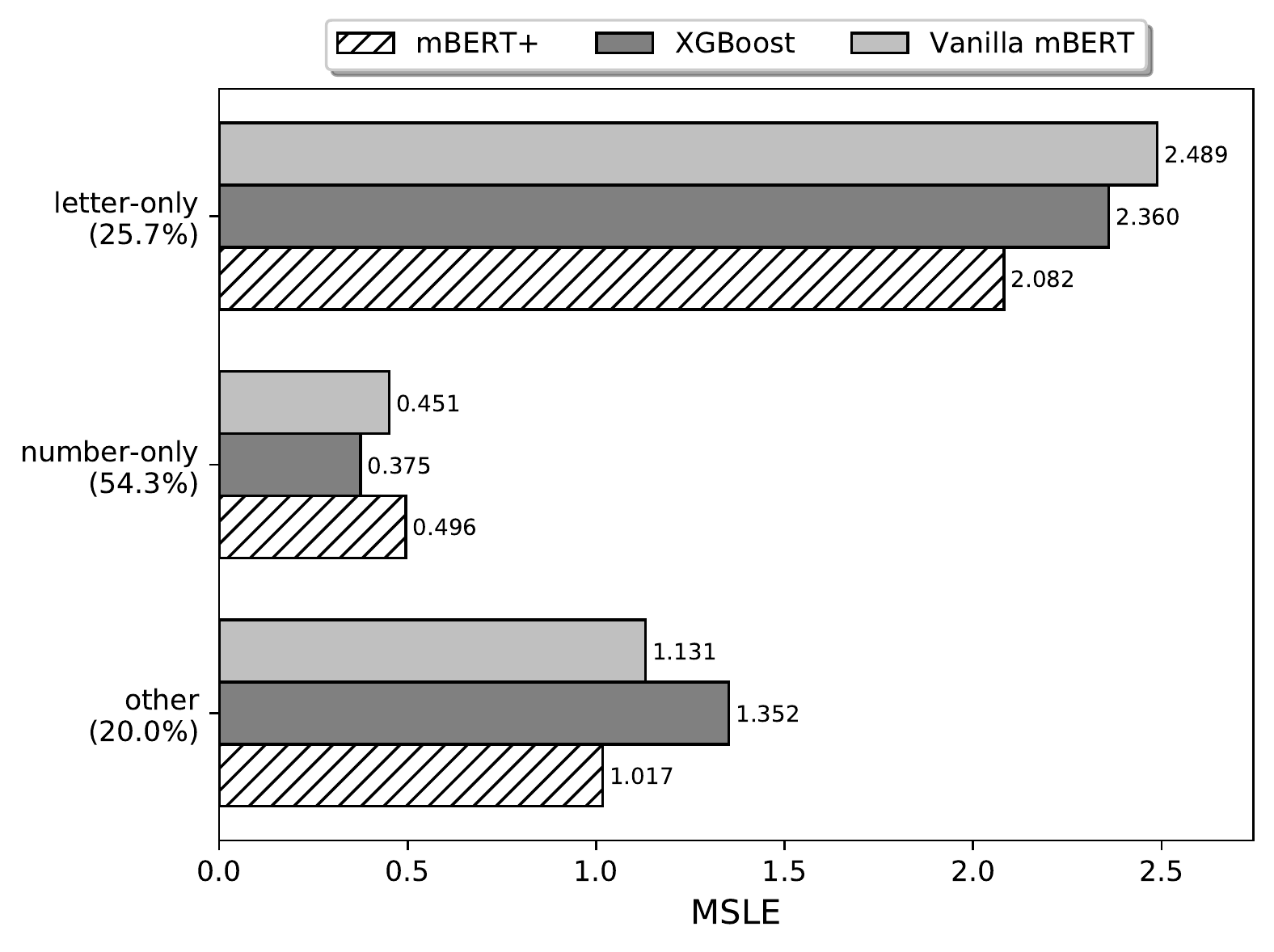}
        \caption{{\dnnft}.}
    \end{subfigure}
    \caption{Performance comparison in MSLE by name character set on the development set. The percentage of each character set group is in parentheses.}
    \label{fig:errorcharset}
\end{figure}

We perform an error analysis of XGBoost, vanilla mBERT, and mBERT$+$ on the development set to understand their difference and identify their limitations.

\subsubsection{Name Length} 

We report the model performance with respect to different name length groups in Figure~\ref{fig:errorlength}. We observe a clear difference in performance over different name length groups for all asset classes. Notably, all models for {\dndomain} and {\dnnft} can give relatively accurate predictions when given an asset of name length four, and all models for {\dnemail} and {\dnnft} perform relatively well when given an asset of name length five. XGBoost demonstrates a notable advantage over the other models in valuing email addresses longer than seven.

\subsubsection{Suffix}

We compare in Figure~\ref{fig:errorsuffix} the model performance for different suffixes. While, unsurprisingly, uncommon suffixes that fall under the ``other'' groups are relatively hard to assess for all asset classes, perhaps surprisingly, models perform worse than average on some common suffixes, including ``net'' in {\dndomain} and ``126.com'' in {\dnemail}. Although mBERT$+$ achieves the lowest MSLE in most groups, XGBoost considerably outperforms mBERT$+$ in a few groups, such as ``other'' in {\dnnft}.

\subsubsection{Name Character Set}

We present the model performance grouped by the character set of the asset name in Figure~\ref{fig:errorcharset}. Interestingly, the relative difficulty in valuing number-only names varies greatly across asset classes. Besides, for all models and all asset classes, the appraisal of letter-only names is more challenging compared with names that fall into the ``other'' groups. XGBoost once again outperforms mBERT$+$ in several groups.

\subsection{Further Discussions}

\subsubsection{Model Ensemble}

Since the error analysis in Section~\ref{sec:error} indicates XGBoost and mBERT$+$ are complementary in many aspects, we combine them by taking the geometric mean of their predictions. As shown in Table~\ref{tab:ensemble}, the ensemble model achieves an MSLE reduction of $5.9\%$ on average compared with mBERT$+$. 
\begin{table}[t!]
\centering
\begin{tabular}{lcc}
\toprule
& \bf Dev & \bf Test \\
\midrule
{\dndomain} & $2.071$ ($\downarrow$0.035) & $1.840$ ($\downarrow$0.052) \\
{\dnemail} & $1.094$ ($\downarrow$0.118) & $1.251$ ($\downarrow$0.117) \\
{\dnnft} & $0.941$ ($\downarrow$0.066) & $0.981$ ($\downarrow$0.085) \\
\midrule
Average & $1.369$ ($\downarrow$0.073) & $1.357$ ($\downarrow$0.085) \\
\bottomrule
\end{tabular}
\caption{Performance of an ensemble of XGBoost and mBERT$+$ in MSLE ($\downarrow$: decreased MSLE compared with mBERT$+$). All MSLE reductions are statistically significant ($\text{all p-values}<5\times 10^{-4}$).} 
\label{tab:ensemble}
\end{table}

\subsubsection{Pre-Trained Language Models} 

We study the impact of different pre-trained language models on the performance of neural models. We choose the following language models for comparison: English uncased BERT-Base~\cite{devlin2019bert} (BERT), XLM-R-Base~\cite{conneau2020unsupervised} (XLM-R), and FNet-Base~\cite{lee2022fnet} (FNet). We denote the vanilla fine-tuned model and the improved model with the pre-trained language model replaced by $\mathcal{M}\in \{\text{BERT}, \text{XLM-R}, \text{FNet}\}$ as vanilla $\mathcal{M}$ and $\mathcal{M}+$, respectively. To minimize time leakage risk, we report the performance on {\dnnft} whose transactions in the development and test sets all happen after the release dates of the employed language models. As shown in Table~\ref{tab:lm}, our improved model consistently outperforms the corresponding vanilla model regardless of the pre-trained language model used. Surprisingly, the relative strength of these pre-trained models differs dramatically between natural language understanding and digital asset valuation: XLM-R does not outperform mBERT in asset valuation, though XLM-R is superior to mBERT in multilingual natural language processing~\cite{conneau2020unsupervised}; similarly, BERT does not outperform FNet, though FNet sacrifices some performance for speed compared with BERT~\cite{lee2022fnet}. Moreover, we find that monolingual pre-trained LMs can achieve performance close to multilingual pre-trained LMs (\eg, FNet \vs mBERT).

\begin{table}[t!]
\centering
\begin{tabular}{lcc}
\toprule
& \bf Dev & \bf Test \\
\midrule
Vanilla mBERT & $1.111$ & $1.312$ \\
Vanilla BERT & $1.193$ & $1.331$ \\
Vanilla XLM-R & $1.147$ & $1.163$ \\
Vanilla FNet & $1.176$ & $1.084$  \\
\midrule
mBERT$+$ & $1.007$ & $1.066$ \\
BERT$+$ & $1.075$ & $1.249$ \\
XLM-R$+$ & $1.058$ & $1.100$ \\
FNet$+$ & $1.046$ & $1.033$ \\
\bottomrule
\end{tabular}
\caption{Performance comparison of models with different pre-trained language models in MSLE on {\dnnft}.} 
\label{tab:lm}
\end{table}

\subsubsection{Non-Uniform Sample Weights}

Motivated by the effectiveness of the proposed two-stage fine-tuning approach, we investigate whether we can improve the conventional model by emphasizing more on the relatively new transactions during training as well. We present the experimental results in Table~\ref{tab:weight}, where we set the sample weights of the newest $T$ transactions to be two times the weights of the other transactions in the training data. We observe no substantial difference in average performance between XGBoost with and without non-uniform sample weights. %
\begin{table}[t!]
\centering
\begin{tabular}{lcc}
\toprule
& \bf Dev & \bf Test \\
\midrule
{\dndomain} & $2.199$ ($\downarrow$0.019) & $1.931$ ($\downarrow$0.009) \\
{\dnemail} & $1.282$ ($\uparrow$0.029) & $1.518$ ($\uparrow$0.021) \\
{\dnnft} & $1.070$ ($\downarrow$0.011) & $1.056$ ($\downarrow$0.021) \\
\midrule
Average & $1.517$ ($=$0.000) & $1.502$ ($\downarrow$0.003) \\
\bottomrule
\end{tabular}
\caption{Performance of XGBoost with non-uniform sample weights ($\uparrow$/$\downarrow$/$=$: increased/decreased/unchanged MSLE compared with XGBoost).} 
\label{tab:weight}
\end{table}

\subsubsection{Comparison with a Commercial Model} 
\label{sec:govalue}

We compare our models with GoValue, a state-of-the-art proprietary automated domain valuation tool from the industry (Section~\ref{sec:related:approach}). Because (i) GoValue only gives the valuation result when the estimated price is between $100$ USD and $25,000$ USD, (ii) GoValue supports valuing neither internationalized domain names (IDNs) nor third-level domain names, and (iii) GoValue does not support bulk appraisal and has a limited query quota, we use a modified setting for the experiment, specified in the following.
\begin{itemize}
    \item We randomly sample from the test set of {\dndomain} $100$ transactions, in which the domain name is neither an IDN nor a third-level domain name, and the price is between $100$ USD and $25,000$ USD.
    \item We call GoValue and our models to predict the price of every domain name in the $100$ samples. If the predicted price is below $100$ (\resp above $25,000$) USD, we change the prediction to $100$ (\resp $25,000$).  %
\end{itemize}

As shown in Table~\ref{tab:govalue}, all our conventional feature-based models and neural models significantly outperform GoValue by a large margin ($\text{all p-values}<1\times 10^{-5}$). Note that although the result is highly suggestive of the superiority of our models, the result may not be conclusive enough because the comparison is arguably unfair for both GoValue and our models. Specifically, on the one hand, GoValue leverages unrivaled amounts of data available only to GoDaddy for model training; on the other hand, the distribution of the test samples is likely closer to that of the training set of {\dndomain} compared with GoValue's training data. Nevertheless, the comparison can hardly be improved due to a lack of access to both the modeling details and training data of GoValue. 

\begin{table}[t!]
\centering
\begin{tabular}{lc}
\toprule
& MSLE \\
\midrule
GoValue & $4.379$ \\
\midrule
AdaBoost & $1.911$ \\
Random Forest & $1.612$ \\
XGBoost & $1.509$ \\
Vanilla mBERT & $1.577$ \\
mBERT$+$ & $1.798$ \\
\bottomrule
\end{tabular}
\caption{Performance comparison using the modified setting (Section~\ref{sec:govalue}) on $100$ samples from {\dndomain}.} 
\label{tab:govalue}
\end{table}

\subsection{Limitations and Future Work}

The data and models presented in this work focus on digital assets represented in texts without touching other modalities (\eg, images). Nevertheless, such a uniform representation makes our work a reasonably suitable starting point for studying general techniques that apply to multiple asset classes. We leave the study of valuing assets represented in other modalities for future work. 

We are aware of some proprietary external knowledge sources (\eg, Google Trends\footnote{\url{https://trends.google.com/}}) that contain potentially additional useful knowledge for building more accurate valuation models. However, we choose not to employ them in this paper to avoid dependence on proprietary business services and leave the study of leveraging them for future research. While this choice limits the best performance our models can attain, we believe it does not influence the main contributions of this paper and helps improve the accessibility and reproducibility of this work.

%% file: 6_conclusion.tex
\section{Conclusion}

We present {\dn}, the first digital asset sales history dataset featuring multiple asset classes, including domain names, email addresses, and NFTs.
We propose several valuation models for {\dn}, including conventional feature-based models and deep learning models, all applicable to multiple asset classes. We conduct comprehensive experiments to evaluate the proposed models on {\dn} and, for the first time, demonstrate that fine-tuning a pre-trained model can beat conventional models in digital asset valuation.

%% file: main.bbl

\begin{thebibliography}{30}


\ifx \showCODEN    \undefined \def \showCODEN     #1{\unskip}     \fi
\ifx \showDOI      \undefined \def \showDOI       #1{#1}\fi
\ifx \showISBNx    \undefined \def \showISBNx     #1{\unskip}     \fi
\ifx \showISBNxiii \undefined \def \showISBNxiii  #1{\unskip}     \fi
\ifx \showISSN     \undefined \def \showISSN      #1{\unskip}     \fi
\ifx \showLCCN     \undefined \def \showLCCN      #1{\unskip}     \fi
\ifx \shownote     \undefined \def \shownote      #1{#1}          \fi
\ifx \showarticletitle \undefined \def \showarticletitle #1{#1}   \fi
\ifx \showURL      \undefined \def \showURL       {\relax}        \fi
\providecommand\bibfield[2]{#2}
\providecommand\bibinfo[2]{#2}
\providecommand\natexlab[1]{#1}
\providecommand\showeprint[2][]{arXiv:#2}

\bibitem[Bikadi et~al\mbox{.}(2017)]%
        {bikadi2017prediction}
\bibfield{author}{\bibinfo{person}{Zsolt Bikadi}, \bibinfo{person}{Sapumal
  Ahangama}, {and} \bibinfo{person}{Eszter Hazai}.}
  \bibinfo{year}{2017}\natexlab{}.
\newblock \showarticletitle{Prediction of Domain Values: High throughput
  screening of domain names using Support Vector Machines}.
\newblock \bibinfo{journal}{\emph{arXiv preprint}}
  \bibinfo{volume}{cs.CY/1707.00906} (\bibinfo{year}{2017}).
\newblock
\urldef\tempurl%
\url{https://doi.org/10.48550/ARXIV.1707.00906}
\showDOI{\tempurl}


\bibitem[Chen and Guestrin(2016)]%
        {chen2016xgboost}
\bibfield{author}{\bibinfo{person}{Tianqi Chen} {and} \bibinfo{person}{Carlos
  Guestrin}.} \bibinfo{year}{2016}\natexlab{}.
\newblock \showarticletitle{XGBoost: A Scalable Tree Boosting System}. In
  \bibinfo{booktitle}{\emph{Proceedings of the 22nd ACM SIGKDD International
  Conference on Knowledge Discovery and Data Mining}} (San Francisco,
  California, USA) \emph{(\bibinfo{series}{KDD '16})}.
  \bibinfo{publisher}{Association for Computing Machinery},
  \bibinfo{address}{New York, NY, USA}, \bibinfo{pages}{785–794}.
\newblock
\showISBNx{9781450342322}
\urldef\tempurl%
\url{https://doi.org/10.1145/2939672.2939785}
\showDOI{\tempurl}


\bibitem[Conneau et~al\mbox{.}(2020)]%
        {conneau2020unsupervised}
\bibfield{author}{\bibinfo{person}{Alexis Conneau}, \bibinfo{person}{Kartikay
  Khandelwal}, \bibinfo{person}{Naman Goyal}, \bibinfo{person}{Vishrav
  Chaudhary}, \bibinfo{person}{Guillaume Wenzek}, \bibinfo{person}{Francisco
  Guzm{\'a}n}, \bibinfo{person}{Edouard Grave}, \bibinfo{person}{Myle Ott},
  \bibinfo{person}{Luke Zettlemoyer}, {and} \bibinfo{person}{Veselin
  Stoyanov}.} \bibinfo{year}{2020}\natexlab{}.
\newblock \showarticletitle{Unsupervised Cross-lingual Representation Learning
  at Scale}. In \bibinfo{booktitle}{\emph{Proceedings of the 58th Annual
  Meeting of the Association for Computational Linguistics}}.
  \bibinfo{publisher}{Association for Computational Linguistics},
  \bibinfo{address}{Online}, \bibinfo{pages}{8440--8451}.
\newblock
\urldef\tempurl%
\url{https://doi.org/10.18653/v1/2020.acl-main.747}
\showDOI{\tempurl}


\bibitem[Del\.{i}ba{\c{s}}(2019)]%
        {delibacs2019domain}
\bibfield{author}{\bibinfo{person}{Emrullah Del\.{i}ba{\c{s}}}.}
  \bibinfo{year}{2019}\natexlab{}.
\newblock \emph{\bibinfo{title}{Domain Name Valuation: Characteristics \& Price
  Exposed!}}
\newblock \bibinfo{thesistype}{Master's\ thesis}. \bibinfo{school}{Istanbul
  {\c{S}}ehir University}.
\newblock


\bibitem[Devlin et~al\mbox{.}(2019)]%
        {devlin2019bert}
\bibfield{author}{\bibinfo{person}{Jacob Devlin}, \bibinfo{person}{Ming-Wei
  Chang}, \bibinfo{person}{Kenton Lee}, {and} \bibinfo{person}{Kristina
  Toutanova}.} \bibinfo{year}{2019}\natexlab{}.
\newblock \showarticletitle{{BERT}: Pre-training of Deep Bidirectional
  Transformers for Language Understanding}. In
  \bibinfo{booktitle}{\emph{Proceedings of the 2019 Conference of the North
  {A}merican Chapter of the Association for Computational Linguistics: Human
  Language Technologies, Volume 1 (Long and Short Papers)}}.
  \bibinfo{publisher}{Association for Computational Linguistics},
  \bibinfo{address}{Minneapolis, Minnesota}, \bibinfo{pages}{4171--4186}.
\newblock
\urldef\tempurl%
\url{https://doi.org/10.18653/v1/N19-1423}
\showDOI{\tempurl}


\bibitem[Dieterle and Bergmann(2014)]%
        {dieterle2014hybrid}
\bibfield{author}{\bibinfo{person}{Sebastian Dieterle} {and}
  \bibinfo{person}{Ralph Bergmann}.} \bibinfo{year}{2014}\natexlab{}.
\newblock \showarticletitle{A Hybrid CBR-ANN Approach to the Appraisal of
  Internet Domain Names}. In \bibinfo{booktitle}{\emph{Case-Based Reasoning
  Research and Development}}, \bibfield{editor}{\bibinfo{person}{Luc
  Lamontagne} {and} \bibinfo{person}{Enric Plaza}} (Eds.).
  \bibinfo{publisher}{Springer International Publishing},
  \bibinfo{address}{Cham}, \bibinfo{pages}{95--109}.
\newblock
\showISBNx{978-3-319-11209-1}
\urldef\tempurl%
\url{https://doi.org/10.1007/978-3-319-11209-1_8}
\showDOI{\tempurl}


\bibitem[Dutta et~al\mbox{.}(2020)]%
        {dutta2020gated}
\bibfield{author}{\bibinfo{person}{Aniruddha Dutta}, \bibinfo{person}{Saket
  Kumar}, {and} \bibinfo{person}{Meheli Basu}.}
  \bibinfo{year}{2020}\natexlab{}.
\newblock \showarticletitle{A Gated Recurrent Unit Approach to Bitcoin Price
  Prediction}.
\newblock \bibinfo{journal}{\emph{Journal of Risk and Financial Management}}
  \bibinfo{volume}{13}, \bibinfo{number}{2} (\bibinfo{year}{2020}).
\newblock
\showISSN{1911-8074}
\urldef\tempurl%
\url{https://doi.org/10.3390/jrfm13020023}
\showDOI{\tempurl}


\bibitem[Freund and Schapire(1995)]%
        {freund1995decision}
\bibfield{author}{\bibinfo{person}{Yoav Freund} {and}
  \bibinfo{person}{Robert~E. Schapire}.} \bibinfo{year}{1995}\natexlab{}.
\newblock \showarticletitle{A desicion-theoretic generalization of on-line
  learning and an application to boosting}. In
  \bibinfo{booktitle}{\emph{Computational Learning Theory}},
  \bibfield{editor}{\bibinfo{person}{Paul Vit{\'a}nyi}} (Ed.).
  \bibinfo{publisher}{Springer Berlin Heidelberg}, \bibinfo{address}{Berlin,
  Heidelberg}, \bibinfo{pages}{23--37}.
\newblock
\showISBNx{978-3-540-49195-8}
\urldef\tempurl%
\url{https://doi.org/10.1007/3-540-59119-2_166}
\showDOI{\tempurl}


\bibitem[Jain et~al\mbox{.}(2022)]%
        {jain2022nft}
\bibfield{author}{\bibinfo{person}{Shrey Jain}, \bibinfo{person}{Camille
  Bruckmann}, {and} \bibinfo{person}{Chase McDougall}.}
  \bibinfo{year}{2022}\natexlab{}.
\newblock \showarticletitle{NFT Appraisal Prediction: Utilizing Search Trends,
  Public Market Data, Linear Regression and Recurrent Neural Networks}.
\newblock \bibinfo{journal}{\emph{arXiv preprint}}
  \bibinfo{volume}{q-fin.ST/2204.12932} (\bibinfo{year}{2022}).
\newblock
\urldef\tempurl%
\url{https://doi.org/10.48550/ARXIV.2204.12932}
\showDOI{\tempurl}


\bibitem[Kapoor et~al\mbox{.}(2022)]%
        {kapoor2022tweetboost}
\bibfield{author}{\bibinfo{person}{Arnav Kapoor}, \bibinfo{person}{Dipanwita
  Guhathakurta}, \bibinfo{person}{Mehul Mathur}, \bibinfo{person}{Rupanshu
  Yadav}, \bibinfo{person}{Manish Gupta}, {and} \bibinfo{person}{Ponnurangam
  Kumaraguru}.} \bibinfo{year}{2022}\natexlab{}.
\newblock \showarticletitle{TweetBoost: Influence of Social Media on NFT
  Valuation}. In \bibinfo{booktitle}{\emph{Companion Proceedings of the Web
  Conference 2022}} (Virtual Event, Lyon, France) \emph{(\bibinfo{series}{WWW
  '22})}. \bibinfo{publisher}{Association for Computing Machinery},
  \bibinfo{address}{New York, NY, USA}, \bibinfo{pages}{621–629}.
\newblock
\showISBNx{9781450391306}
\urldef\tempurl%
\url{https://doi.org/10.1145/3487553.3524642}
\showDOI{\tempurl}


\bibitem[Lee-Thorp et~al\mbox{.}(2022)]%
        {lee2022fnet}
\bibfield{author}{\bibinfo{person}{James Lee-Thorp}, \bibinfo{person}{Joshua
  Ainslie}, \bibinfo{person}{Ilya Eckstein}, {and} \bibinfo{person}{Santiago
  Ontanon}.} \bibinfo{year}{2022}\natexlab{}.
\newblock \showarticletitle{{FN}et: Mixing Tokens with {F}ourier Transforms}.
  In \bibinfo{booktitle}{\emph{Proceedings of the 2022 Conference of the North
  American Chapter of the Association for Computational Linguistics: Human
  Language Technologies}}. \bibinfo{publisher}{Association for Computational
  Linguistics}, \bibinfo{address}{Seattle, United States},
  \bibinfo{pages}{4296--4313}.
\newblock
\urldef\tempurl%
\url{https://doi.org/10.18653/v1/2022.naacl-main.319}
\showDOI{\tempurl}


\bibitem[Lindenthal(2014)]%
        {lindenthal2014valuable}
\bibfield{author}{\bibinfo{person}{Thies Lindenthal}.}
  \bibinfo{year}{2014}\natexlab{}.
\newblock \showarticletitle{Valuable words: The price dynamics of internet
  domain names}.
\newblock \bibinfo{journal}{\emph{Journal of the Association for Information
  Science and Technology}} \bibinfo{volume}{65}, \bibinfo{number}{5}
  (\bibinfo{year}{2014}), \bibinfo{pages}{869--881}.
\newblock
\urldef\tempurl%
\url{https://doi.org/10.1002/asi.23012}
\showDOI{\tempurl}


\bibitem[Liu et~al\mbox{.}(2019)]%
        {liu2019data}
\bibfield{author}{\bibinfo{person}{Jian Liu}, \bibinfo{person}{Xiangdong Zeng},
  \bibinfo{person}{Adam Ghandar}, {and} \bibinfo{person}{Georgios
  Theodoropoulos}.} \bibinfo{year}{2019}\natexlab{}.
\newblock \showarticletitle{Data Driven Domain Appraisal: Extracting
  Information from Short Dense Texts}. In \bibinfo{booktitle}{\emph{2019 IEEE
  Symposium Series on Computational Intelligence (SSCI)}}.
  \bibinfo{pages}{2489--2496}.
\newblock
\urldef\tempurl%
\url{https://doi.org/10.1109/SSCI44817.2019.9002904}
\showDOI{\tempurl}


\bibitem[Mekacher et~al\mbox{.}(2022)]%
        {mekacher2022heterogeneous}
\bibfield{author}{\bibinfo{person}{Amin Mekacher}, \bibinfo{person}{Alberto
  Bracci}, \bibinfo{person}{Matthieu Nadini}, \bibinfo{person}{Mauro Martino},
  \bibinfo{person}{Laura Alessandretti}, \bibinfo{person}{Luca~Maria Aiello},
  {and} \bibinfo{person}{Andrea Baronchelli}.} \bibinfo{year}{2022}\natexlab{}.
\newblock \showarticletitle{Heterogeneous rarity patterns drive price dynamics
  in NFT collections}.
\newblock \bibinfo{journal}{\emph{Scientific Reports}} \bibinfo{volume}{12},
  \bibinfo{number}{1} (\bibinfo{date}{16 Aug} \bibinfo{year}{2022}),
  \bibinfo{pages}{13890}.
\newblock
\showISSN{2045-2322}
\urldef\tempurl%
\url{https://doi.org/10.1038/s41598-022-17922-5}
\showDOI{\tempurl}


\bibitem[Meystedt(2015)]%
        {meystedt2015my}
\bibfield{author}{\bibinfo{person}{Aron Meystedt}.}
  \bibinfo{year}{2015}\natexlab{}.
\newblock \showarticletitle{What is my URL worth? Placing a value on premium
  domain names}.
\newblock \bibinfo{journal}{\emph{Valuation Strategies}} \bibinfo{volume}{19},
  \bibinfo{number}{2} (\bibinfo{year}{2015}), \bibinfo{pages}{10--17,48}.
\newblock


\bibitem[Miramirkhani et~al\mbox{.}(2018)]%
        {miramirkhani2018panning}
\bibfield{author}{\bibinfo{person}{Najmeh Miramirkhani},
  \bibinfo{person}{Timothy Barron}, \bibinfo{person}{Michael Ferdman}, {and}
  \bibinfo{person}{Nick Nikiforakis}.} \bibinfo{year}{2018}\natexlab{}.
\newblock \showarticletitle{Panning for Gold.Com: Understanding the Dynamics of
  Domain Dropcatching}. In \bibinfo{booktitle}{\emph{Proceedings of the 2018
  World Wide Web Conference}} (Lyon, France) \emph{(\bibinfo{series}{WWW
  '18})}. \bibinfo{publisher}{International World Wide Web Conferences Steering
  Committee}, \bibinfo{address}{Republic and Canton of Geneva, CHE},
  \bibinfo{pages}{257--266}.
\newblock
\showISBNx{9781450356398}
\urldef\tempurl%
\url{https://doi.org/10.1145/3178876.3186092}
\showDOI{\tempurl}


\bibitem[Nadini et~al\mbox{.}(2021)]%
        {nadini2021mapping}
\bibfield{author}{\bibinfo{person}{Matthieu Nadini}, \bibinfo{person}{Laura
  Alessandretti}, \bibinfo{person}{Flavio Di~Giacinto}, \bibinfo{person}{Mauro
  Martino}, \bibinfo{person}{Luca~Maria Aiello}, {and} \bibinfo{person}{Andrea
  Baronchelli}.} \bibinfo{year}{2021}\natexlab{}.
\newblock \showarticletitle{Mapping the NFT revolution: market trends, trade
  networks, and visual features}.
\newblock \bibinfo{journal}{\emph{Scientific Reports}} \bibinfo{volume}{11},
  \bibinfo{number}{1} (\bibinfo{date}{22 Oct} \bibinfo{year}{2021}),
  \bibinfo{pages}{20902}.
\newblock
\showISSN{2045-2322}
\urldef\tempurl%
\url{https://doi.org/10.1038/s41598-021-00053-8}
\showDOI{\tempurl}


\bibitem[Nguyen(2001)]%
        {nguyen2001cyberproperty}
\bibfield{author}{\bibinfo{person}{Xuan-Thao Nguyen}.}
  \bibinfo{year}{2001}\natexlab{}.
\newblock \showarticletitle{Cyberproperty and Judicial Dissonance: The Trouble
  with Domain Name Classification}.
\newblock \bibinfo{journal}{\emph{George Mason Law Review}}
  \bibinfo{volume}{10}, \bibinfo{number}{2} (\bibinfo{year}{2001}),
  \bibinfo{pages}{183--214}.
\newblock


\bibitem[Pedregosa et~al\mbox{.}(2011)]%
        {pedregosa2011scikit}
\bibfield{author}{\bibinfo{person}{Fabian Pedregosa},
  \bibinfo{person}{Ga{{\"e}}l Varoquaux}, \bibinfo{person}{Alexandre Gramfort},
  \bibinfo{person}{Vincent Michel}, \bibinfo{person}{Bertrand Thirion},
  \bibinfo{person}{Olivier Grisel}, \bibinfo{person}{Mathieu Blondel},
  \bibinfo{person}{Peter Prettenhofer}, \bibinfo{person}{Ron Weiss},
  \bibinfo{person}{Vincent Dubourg}, \bibinfo{person}{Jake Vanderplas},
  \bibinfo{person}{Alexandre Passos}, \bibinfo{person}{David Cournapeau},
  \bibinfo{person}{Matthieu Brucher}, \bibinfo{person}{Matthieu Perrot}, {and}
  \bibinfo{person}{{{\'E}}douard Duchesnay}.} \bibinfo{year}{2011}\natexlab{}.
\newblock \showarticletitle{Scikit-learn: Machine Learning in Python}.
\newblock \bibinfo{journal}{\emph{Journal of Machine Learning Research}}
  \bibinfo{volume}{12}, \bibinfo{number}{85} (\bibinfo{year}{2011}),
  \bibinfo{pages}{2825--2830}.
\newblock


\bibitem[Pennington et~al\mbox{.}(2014)]%
        {pennington2014glove}
\bibfield{author}{\bibinfo{person}{Jeffrey Pennington},
  \bibinfo{person}{Richard Socher}, {and} \bibinfo{person}{Christopher
  Manning}.} \bibinfo{year}{2014}\natexlab{}.
\newblock \showarticletitle{{G}lo{V}e: Global Vectors for Word Representation}.
  In \bibinfo{booktitle}{\emph{Proceedings of the 2014 Conference on Empirical
  Methods in Natural Language Processing ({EMNLP})}}.
  \bibinfo{publisher}{Association for Computational Linguistics},
  \bibinfo{address}{Doha, Qatar}, \bibinfo{pages}{1532--1543}.
\newblock
\urldef\tempurl%
\url{https://doi.org/10.3115/v1/D14-1162}
\showDOI{\tempurl}


\bibitem[Polu and Sutskever(2020)]%
        {polu2020generative}
\bibfield{author}{\bibinfo{person}{Stanislas Polu} {and} \bibinfo{person}{Ilya
  Sutskever}.} \bibinfo{year}{2020}\natexlab{}.
\newblock \showarticletitle{Generative Language Modeling for Automated Theorem
  Proving}.
\newblock \bibinfo{journal}{\emph{arXiv preprint}}
  \bibinfo{volume}{cs.LG/2009.03393} (\bibinfo{year}{2020}).
\newblock
\urldef\tempurl%
\url{https://doi.org/10.48550/ARXIV.2009.03393}
\showDOI{\tempurl}


\bibitem[Radford et~al\mbox{.}(2018)]%
        {radford2018improving}
\bibfield{author}{\bibinfo{person}{Alec Radford}, \bibinfo{person}{Karthik
  Narasimhan}, \bibinfo{person}{Tim Salimans}, {and} \bibinfo{person}{Ilya
  Sutskever}.} \bibinfo{year}{2018}\natexlab{}.
\newblock \bibinfo{booktitle}{\emph{Improving Language Understanding by
  Generative Pre-Training}}.
\newblock \bibinfo{type}{{T}echnical {R}eport}.
\newblock


\bibitem[St{\"o}ckl(2021)]%
        {stockl2021watching}
\bibfield{author}{\bibinfo{person}{Andreas St{\"o}ckl}.}
  \bibinfo{year}{2021}\natexlab{}.
\newblock \showarticletitle{Watching a Language Model Learning Chess}. In
  \bibinfo{booktitle}{\emph{Proceedings of the International Conference on
  Recent Advances in Natural Language Processing (RANLP 2021)}}.
  \bibinfo{publisher}{INCOMA Ltd.}, \bibinfo{address}{Held Online},
  \bibinfo{pages}{1369--1379}.
\newblock


\bibitem[Tang et~al\mbox{.}(2014)]%
        {tang2014general}
\bibfield{author}{\bibinfo{person}{{Jih Hsin} Tang}, \bibinfo{person}{{Min Chu}
  Hsu}, \bibinfo{person}{{Ting Yuan} Hu}, {and} \bibinfo{person}{Hsing-Hua
  Huang}.} \bibinfo{year}{2014}\natexlab{}.
\newblock \showarticletitle{A general domain name appraisal model}.
\newblock \bibinfo{journal}{\emph{Journal of Internet Technology}}
  \bibinfo{volume}{15}, \bibinfo{number}{3} (\bibinfo{date}{1 May}
  \bibinfo{year}{2014}), \bibinfo{pages}{427--431}.
\newblock
\showISSN{1607-9264}
\urldef\tempurl%
\url{https://doi.org/10.6138/JIT.2014.15.3.11}
\showDOI{\tempurl}


\bibitem[Virpioja et~al\mbox{.}(2013)]%
        {virpioja2013morfessor}
\bibfield{author}{\bibinfo{person}{Sami Virpioja}, \bibinfo{person}{Peter
  Smit}, \bibinfo{person}{Stig-Arne Grönroos}, {and} \bibinfo{person}{Mikko
  Kurimo}.} \bibinfo{year}{2013}\natexlab{}.
\newblock \bibinfo{booktitle}{\emph{{Morfessor 2.0: Python Implementation and
  Extensions for Morfessor Baseline}}}.
\newblock \bibinfo{type}{{T}echnical {R}eport}. \bibinfo{pages}{38} pages.
\newblock
\showISBNx{978-952-60-5501-5 (electronic)}
\showISSN{1799-490X (electronic), 1799-4896 (printed), 1799-4896 (ISSN-L)}


\bibitem[Visconti(2020)]%
        {visconti2020valuation}
\bibfield{author}{\bibinfo{person}{Roberto~Moro Visconti}.}
  \bibinfo{year}{2020}\natexlab{}.
\newblock \bibinfo{booktitle}{\emph{The Valuation of Digital Intangibles:
  Technology, Marketing and Internet}}.
\newblock \bibinfo{publisher}{Palgrave Macmillan Cham}.
\newblock
\showISBNx{9783030369187}
\urldef\tempurl%
\url{https://doi.org/10.1007/978-3-030-36918-7}
\showDOI{\tempurl}


\bibitem[Wolf et~al\mbox{.}(2020)]%
        {wolf2020transformers}
\bibfield{author}{\bibinfo{person}{Thomas Wolf}, \bibinfo{person}{Lysandre
  Debut}, \bibinfo{person}{Victor Sanh}, \bibinfo{person}{Julien Chaumond},
  \bibinfo{person}{Clement Delangue}, \bibinfo{person}{Anthony Moi},
  \bibinfo{person}{Pierric Cistac}, \bibinfo{person}{Tim Rault},
  \bibinfo{person}{Remi Louf}, \bibinfo{person}{Morgan Funtowicz},
  \bibinfo{person}{Joe Davison}, \bibinfo{person}{Sam Shleifer},
  \bibinfo{person}{Patrick von Platen}, \bibinfo{person}{Clara Ma},
  \bibinfo{person}{Yacine Jernite}, \bibinfo{person}{Julien Plu},
  \bibinfo{person}{Canwen Xu}, \bibinfo{person}{Teven Le~Scao},
  \bibinfo{person}{Sylvain Gugger}, \bibinfo{person}{Mariama Drame},
  \bibinfo{person}{Quentin Lhoest}, {and} \bibinfo{person}{Alexander Rush}.}
  \bibinfo{year}{2020}\natexlab{}.
\newblock \showarticletitle{Transformers: State-of-the-Art Natural Language
  Processing}. In \bibinfo{booktitle}{\emph{Proceedings of the 2020 Conference
  on Empirical Methods in Natural Language Processing: System Demonstrations}}.
  \bibinfo{publisher}{Association for Computational Linguistics},
  \bibinfo{address}{Online}, \bibinfo{pages}{38--45}.
\newblock
\urldef\tempurl%
\url{https://doi.org/10.18653/v1/2020.emnlp-demos.6}
\showDOI{\tempurl}


\bibitem[Wu and He(2009)]%
        {wu2009domain2}
\bibfield{author}{\bibinfo{person}{Zu-guang Wu} {and} \bibinfo{person}{Hai-yi
  He}.} \bibinfo{year}{2009}\natexlab{}.
\newblock \showarticletitle{Domain Name Valuation Model Constructing and
  Emperical Evidence}. In \bibinfo{booktitle}{\emph{2009 International
  Conference on Multimedia Information Networking and Security}},
  Vol.~\bibinfo{volume}{2}. \bibinfo{pages}{201--204}.
\newblock
\urldef\tempurl%
\url{https://doi.org/10.1109/MINES.2009.153}
\showDOI{\tempurl}


\bibitem[Wu et~al\mbox{.}(2009)]%
        {wu2009domain}
\bibfield{author}{\bibinfo{person}{Zu-guang Wu}, \bibinfo{person}{Guo-hua Zhu},
  \bibinfo{person}{Rui Huang}, {and} \bibinfo{person}{Bin Xia}.}
  \bibinfo{year}{2009}\natexlab{}.
\newblock \showarticletitle{Domain Name Valuation Model Based on Semantic
  Theory and Content Analysis}. In \bibinfo{booktitle}{\emph{2009 Asia-Pacific
  Conference on Information Processing}}, Vol.~\bibinfo{volume}{2}.
  \bibinfo{pages}{237--240}.
\newblock
\urldef\tempurl%
\url{https://doi.org/10.1109/APCIP.2009.194}
\showDOI{\tempurl}


\bibitem[Xia et~al\mbox{.}(2021)]%
        {xia2021ethereum}
\bibfield{author}{\bibinfo{person}{Pengcheng Xia}, \bibinfo{person}{Haoyu
  Wang}, \bibinfo{person}{Zhou Yu}, \bibinfo{person}{Xinyu Liu},
  \bibinfo{person}{Xiapu Luo}, {and} \bibinfo{person}{Guoai Xu}.}
  \bibinfo{year}{2021}\natexlab{}.
\newblock \showarticletitle{Ethereum Name Service: the Good, the Bad, and the
  Ugly}.
\newblock \bibinfo{journal}{\emph{arXiv preprint}}
  \bibinfo{volume}{cs.CR/2104.05185} (\bibinfo{year}{2021}).
\newblock
\urldef\tempurl%
\url{https://doi.org/10.48550/ARXIV.2104.05185}
\showDOI{\tempurl}


\end{thebibliography}
